\documentclass[aps,floatfix,amsmath,nofootinbib,amssymb,superscriptaddress,letterpaper,twocolumn]{revtex4-2}

\usepackage{overpic}
\usepackage{amssymb}
\usepackage{indentfirst}
\usepackage{feynmf}
\usepackage{slashed}
\usepackage{cases}
\usepackage{color}
\usepackage{multirow}
\usepackage{epstopdf}
\usepackage{graphicx,color,bm}
\usepackage{booktabs}
\usepackage{float}
\usepackage{amsmath}
\allowdisplaybreaks[4]

\newcommand{\kt}{\bm{k}_T^2}
\newcommand{\ktquad}{\bm{k}_T^4}

\usepackage[colorlinks=true,
            citecolor=green,
            anchorcolor=red,
            menucolor=red,
            linkcolor=red,
            filecolor=red,
            runcolor=red,
            urlcolor=blue]{hyperref}

\begin{document}

\title{Twist-3 transverse momentum dependent gluon distributions in a spectator model}

\author{Xiupeng Xie}\affiliation{School of Physics, Southeast University, Nanjing
211189, China}

\author{Zhun Lu}
\email{zhunlu@seu.edu.cn}
\affiliation{School of Physics, Southeast University, Nanjing 211189, China}

\begin{abstract}
We study the twist-3 transverse momentum dependent distributions of gluons in a nucleon within a spectator model framework. In this model, the nucleon is described as emitting a virtual (time-like) gluon and an on-shell spectator particle, with the spectator mass treated continuously via a spectral function. The twist-3 gluon-gluon correlators, $\Phi^{+i;+-}$ and $\Phi^{ij;l+}$, are parameterized by a set of complex functions, whose real and imaginary parts correspond to T-even and T-odd components, respectively. We numerically check the equation of motion relation for these distributions and find that relation holds fairly well in the spectator model.

\end{abstract}

\maketitle

\section{Introduction}
Over the past few decades, twist-3 effects~\cite{Efremov:1981sh,Qiu:1991wg,Qiu:1998ia}, a key class of subleading-power contributions in quantum chromodynamics (QCD), have attracted considerable theoretical and phenomenological interest. While subleading terms are suppressed by powers of $1/Q$, where $Q$ denotes the hard momentum transfer scale defining the underlying hard-scattering process, the magnitude of twist-3 transverse momentum dependent parton distributions (TMDs) can match that of leading-twist (twist-2) TMDs in specific kinematic regimes, most notably when $Q$ is not asymptotically large. A rigorous understanding of twist-3 TMDs is essential for a complete description of semi-inclusive hard-scattering reactions. Furthermore, these distributions are indispensable for the unbiased high-precision extraction of leading-power quantities from experimental measurements.

In contrast to the well established theoretical framework for leading-twist TMD observables, the theoretical description of subleading-power TMD observables remains an open challenge compared to the current state-of-the-art for leading power observables. Several theoretical methods and corresponding twist-3 cross section formula have been established for various processes. These include the $g_2$-structure function of the nucleon measured in deep-inelastic scattering~\cite{Jaffe:1989xx,Belitsky:2000pb}, single spin asymmetries (SSAs) for a hadron or photon production in proton-proton (nucleus) collisions~\cite{Qiu:1991wg,Qiu:1998ia,Kanazawa:2000hz,Kanazawa:2000kp,Ji:2006vf,Kouvaris:2006zy,Koike:2007rq,
Koike:2009ge,Vogelsang:2009pj,Koike:2011mb,Koike:2011nx,Kang:2010zzb,Kanazawa:2011er,Metz:2012ct,Beppu:2013uda,
Kanazawa:2014nea,Kang:2011ni,Hatta:2016wjz,Hatta:2016khv,Benic:2018moa,Benic:2018amn} and SIDIS~\cite{Ji:2006br,Eguchi:2006qz,Eguchi:2006mc,Yuan:2009dw,Beppu:2010qn,Koike:2011ns,Kanazawa:2013uia,
Yoshida:2016tfh,Xing:2019ovj,Benic:2019zvg}; SSA in transversely polarized hyperon production in the unpolarized proton-proton collision~\cite{Kanazawa:2000cx,Zhou:2008fb,Koike:2015zya,Koike:2017fxr,Yabe:2019awq,Kenta:2019bxd} and in $e^+ e^-$ collision~\cite{Gamberg:2018fwy}; and the longitudinal-transverse double spin asymmetry $A_{LT}$ in the proton-proton collision~\cite{Jaffe:1991ra,Liang:2012rb,Hatta:2013wsa,Koike:2015yza,Koike:2016ura}, etc.

Unlike collinear twist-3 parton distribution functions (PDFs), which have been extensively studied and are typically categorized as intrinsic, kinematical, or dynamical~\cite{Kanazawa:2015ajw} ones, twist-3 transverse-momentum-dependent distributions (TMDs) are subject to stringent constraints from the QCD equations of motion and Lorentz invariance. Consequently, most subleading-power TMDs are not independent to each other. In the quark sector, the quark-gluon-quark (qgq) correlators~\cite{Bacchetta:2004zf}--which encode dynamical power corrections--represent the only genuine source of independent subleading-twist TMDs. All other $1/Q$-suppressed contributions can be systematically expressed in terms of leading-twist TMDs~\cite{Bacchetta:2006tn,Gamberg:2022lju,Ebert:2021jhy}.

In this paper, we study twist-3 TMDs of gluons in a nucleon within a spectator model framework. The central premise of the spectator model is that the nucleon emits a time-like off-shell gluon, together with a single on-shell spectator particle. The spectator mass can take real values in a continuous range described by a spectral function. We model the nucleon-gluon-spectator vertex in analogy to the conserved electromagnetic current of a free nucleon obtained from the Gordon decomposition. 

We identify a complete set of eight twist-3 gluon TMDs. These distributions arise from the gluon-gluon correlators $\Phi^{+i;+-}$ and $\Phi^{ij;l+}$, while the correlator $\Phi^{+i;+j}$ corresponds to the twist-2 case. The twist-3 gluon TMDs are formulated as complex functions, where the T-even functions correspond to the real parts and the T-odd functions correspond to the imaginary parts. Their calculation requires evaluation of the correlators at both tree-level and one-loop level. At tree level, we obtain the real parts (T-even) of twist-3 gluon TMDs. At one-loop level, two distinct types of imaginary parts (T-odd) emerge: the Weizs$\ddot{a}$cker-Williams (WW) gluon TMDs (or $f$-type), and the dipole gluon TMDs (or $d$-type). The two classes are distinguished by the different Wilson line paths in their operator definition. 

Upon integration over the gluon transverse momentum, only two twist-3 collinear PDFs survive: $\Delta G_{3T}\left(x\right)$ and $\Delta H_{3T}\left(x\right)$, which are related to the kinematical PDF $\Delta G_T^{(1)}\left(x\right)$ and the dynamical PDFs via two equations of motion relation~\cite{Koike:2019zxc}. We numerically verify these equations of motion relation within our spectator model framework and find that they are satisfied to excellent numerical accuracy.

The rest of the paper is organized as follows. In Sec.~\ref{section2}, we present the theoretical formalism of the spectator model for twist-3 gluon TMDs. We derive the model expressions for the gluon-gluon correlators and the corresponding twist-3 gluon TMDs using appropriate projecting operators, at both tree-level and one-loop level. Furthermore, we calculate the three-gluon correlator within the same framework to obtain the dynamical twist-3 collinear PDFs, which enables us to validate the aforementioned equation of motion relation. In Sec.~\ref{section3}, we present numerical predictions for the full set of twist-3 gluon TMDs. We additionally show results for the integrated twist-3 TMDs and the dynamical collinear PDFs. Finally, we summarize our key results and conclusions in Sec.~\ref{section4}.

\section{Analytic calculation of the twist-3 TMDs}\label{section2}

Leading and non-leading contributions to the hadronic tensor can be systematically distinguished via a light-cone decomposition of the hadron momentum and spin vectors, defined with respect to two light-like directions $n_+ \equiv \left[1,0,\bm{0}_T\right]$ and $n_-\equiv \left[0,1,\bm{0}_T\right]$. A general light-cone vector $a^\mu$ can be expressed as $\left[a^+, a^-, \bm{a}_T\right]$ or equivalently as $a^+ n_+^\mu +a^- n_-^\mu +a_T^\mu$, where $a_T^\mu \equiv \bm{a}_T$ denotes the transverse component. In a reference frame in which the hadron has no transverse momentum, the hadron momentum $P$, the parton momentum $k$ and the spin vector of the hadron are given by:
\begin{align}
P^\mu=&P^+ n_+^\mu + \frac{M^2}{2P^+} n_-^\mu \,,\\
k^\mu=&xP^+ n_+^\mu + \frac{k^2+\kt }{2xP^+} n_-^\mu +\bm{k}_T \,,\\
S^\mu=&S_L \frac{P^+}{M} n_+^\mu-S_L\frac{M}{2P^+} n_-^\mu+ S_T^\mu\,,
\end{align}
where $M$ is the hadron mass and $x=k^+/P^+$ is the longitudinal momentum fraction of the hadron carried by the parton. The gauge-invariant gluon-gluon correlator is expressed as~\cite{Mulders:2000sh}:
\begin{align}
\Phi&^{\mu \nu;\rho \sigma}\left(x,\bm{k}_T;S\right)=\frac{1}{x}\int \frac{d\xi^- d \bm{\xi}_T}{(2\pi)^3} e^{ik\cdot \xi} \notag\\
&\times \left. \left\langle P,S\left|F_a^{\mu \nu}\left(0\right) \mathcal{U}_{ab}\left(0,\xi\right) F_b^{\rho \sigma}\left(\xi\right)  \right|P,S\right\rangle\right|_{\xi^+=0}\,.\label{eq:phi0}
\end{align}
The gluon field strength tensor $F_a^{\mu \nu}$ is related to the gluon gauge field $A_a^\mu$ by $F_a^{\mu \nu}=\partial^\mu A_a^\nu-\partial^\nu A_a^\mu+g f_{abc} A_b^\mu A_c^\nu$, where $f_{abc}$ is the structure constants of the SU(3) color gauge group and $g$ is the strong coupling constant. The symbol $\mathcal{U}_{ab}\left(0,\xi\right)$ denotes the gauge-link operator
\begin{align}
\mathcal{U}_{ab}\left(0,\xi\right)=\mathcal{P} exp\left[-gf_{abc}\int_{0}^{\xi} d\omega \cdot A_c(\omega)\right]\,.
\end{align}
A gauge link with color flowing through a future-pointing closed Wilson line ($[+,+]$) corresponds to final-state interactions between the spectator and an outgoing gluon. Conversely, initial-state interactions are described by past-pointing closed Wilson lines ($[-,-]$). The gluon TMDs arising from these gauge links are referred to as Weizs$\ddot{a}$cker-Williams (WW) gluon TMDs, or $f$-type gluon TMDs, because the color structure of the T-odd ones involves the antisymmetric structure constants $f_{abc}$ of the color SU(3) group. On the other hand, color can also flow through a closed path involving both initial and final states ($[+,-]$ and $[-,+]$). The corresponding gluon TMDs are usually called dipole gluon TMDs, or $d$-type gluon TMDs, which involve the symmetric structure constants $d_{abc}$ of the color gauge group SU(3). By definition, T-even gluon TMDs are symmetric under interchange of the Wilson line directions, while the T-odd ones change sign under this operation. For the T-even unpolarized function $f_1$, and the T-odd gluon Sivers function $f_{1T}^\perp$, this behavior is encoded in the following modified universality relations~\cite{Boer:2016fqd,Bomhof:2006dp,Buffing:2013kca}:
\begin{align}
f_1^{[+,+]}=&f_1^{[-,-]}\,, &\quad f_1^{[+,-]}&=f_1^{[-,+]}\,,\label{eq:gauge1} \\
f_{1T}^{\perp[+,+]}=&-f_{1T}^{\perp[-,-]}\,, &\quad f_{1T}^{\perp[+,-]}&=-f_{1T}^{\perp[-,+]}\,,\label{eq:gauge2}
\end{align}
It should be emphasized that $f$-type and $d$-type gluon TMDs are physically distinct observables and cannot be directly related; they encode different aspects of gluonic information.

\subsection{The real parts of twist-3 TMDs}

Owing to the antisymmetry of the gluon field strength tensor in Eq.~\eqref{eq:phi0}, the correlator $\Phi^{+i;+j}$ refers to as the twist-2 sector, while $\Phi^{+i;+-} \equiv \Phi^{i-}$ and $\Phi^{ij;l+} \equiv \Phi^{ij;l}$ refer to as the twist-3 sector, where $i,j,l,...$ denote transverse indices. Each of these correlators is parametrized in terms of the corresponding twist-3 gluon TMDs. 
For unpolarized ($U$), longitudinally polarized ($L$), and transversely polarized ($T$) nucleons, we have the following explicit expressions~\cite{Mulders:2000sh},
\begin{align}
\Phi_U^{i-}\left(x,\bm{k}_T\right)=&\frac{k_T^i}{2} G_3^\perp \left(x,\kt \right)\,,\label{eq:phi1} \\
\Phi_L^{i-}\left(x,\bm{k}_T\right)=&\frac{1}{2} i S_L \epsilon_T^{k_T i} \Delta G_{3L}^\perp \left(x,\kt \right)\,,\\
\Phi_T^{i-}\left(x,\bm{k}_T\right)=&\frac{1}{2} \bigg[iM \epsilon_T^{S_T i} \Delta G_{3T}^\prime \left(x,\kt \right)\notag\\
&+ i \epsilon_T^{k_T i} \frac{\bm{k}_T \cdot \bm{S}_T}{M} \Delta G_{3T}^\perp \left(x,\kt \right)\bigg]\notag\\
=&\frac{1}{2} \bigg[iM \epsilon_T^{S_T i} \Delta G_{3T} \left(x,\kt \right)\notag\\
&+i \epsilon_T^{i\alpha}S_T^{\beta} \bigg(\frac{k_{T\alpha} k_{T\beta}}{M} + g_{T\alpha \beta} \frac{\kt }{2M}\bigg)\notag\\
&\times \Delta G_{3T}^\perp \left(x,\kt \right)\bigg]\,,
\end{align}
and
\begin{align}
\Phi_U^{ij;l}\left(x,\bm{k}_T\right)=&\frac{-g_T^{l[i} k_T^{j]}}{2}  H_3^\perp \left(x,\kt \right)\,,\\
\Phi_L^{ij;l}\left(x,\bm{k}_T\right)=&\frac{1}{2} i S_L \epsilon_T^{ij} k_T^l \Delta H_{3L}^\perp \left(x,\kt \right)\,,\\
\Phi_T^{ij;l}\left(x,\bm{k}_T\right)=&\frac{1}{2} \bigg[iM \epsilon_T^{ij} S_T^l \Delta H_{3T}^\prime \left(x,\kt \right)\notag\\
&+ i \epsilon_T^{ij} k_T^l \frac{\bm{k}_T \cdot \bm{S}_T}{M} \Delta H_{3T}^\perp \left(x,\kt \right)\bigg]\notag\\
=&\frac{1}{2} \bigg[iM \epsilon_T^{ij} S_T^l \Delta H_{3T} \left(x,\kt \right)\notag\\
&-i \epsilon_T^{ij}S_{T\alpha} \bigg(\frac{k_{T}^\alpha k_{T}^l}{M} + g_{T}^{\alpha l} \frac{\kt }{2M}\bigg)\notag\\
&\times \Delta H_{3T}^\perp \left(x,\kt \right)\bigg]\,,\label{eq:phi2}
\end{align}
where $g_T^{ij}=g^{ij}-n_+^i n_-^j-n_+^j n_-^i$ and $\epsilon_T^{ij}=\epsilon^{n_+ n_- ij}=\epsilon^{-+ij}$ are the symmetric and antisymmetric transverse tensors, respectively. 
The sign convention is chosen such that $\epsilon_T^{12}=1$. 
While twist-2 TMDs are real functions, twist-3 TMDs are arranged in terms of complex functions in such a way that the T-even functions correspond to the real parts and the T-odd functions correspond to the imaginary parts. 

\begin{figure}[htbp]
    \centering
    \includegraphics[width=0.8\columnwidth]{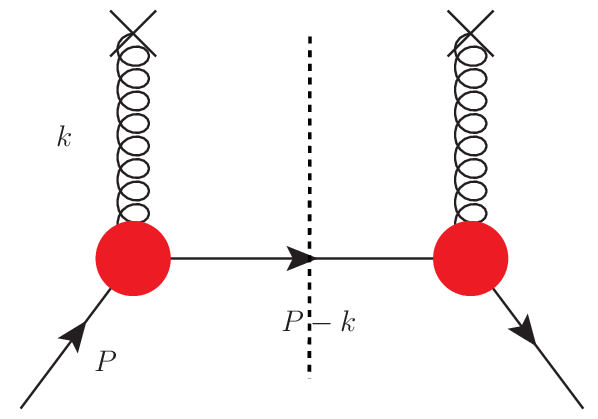}
    \caption{Tree-level cut diagram for the calculation of real parts of Twist-3 gluon TMDs. The red blob represents the nucleon-gluon-spectator vertex. Gluon lines with crosses correspond to specific Feynman rules for the gluon field strength tensor.}\label{fig:tree-level}     
\end{figure}

In this section, we restrict our calculation to leading-order contributions, neglecting the effect of the gauge link and its process dependence. The real parts of the twist-3 gluon TMDs are derived from the tree-level cut diagram shown in Fig.~\ref{fig:tree-level}. We perform the calculation within the spectator approximation, in which the nucleon in the state $|P,S\rangle$ can split into a gluon with momentum $k$ and a single on-shell spectator particle in the state $|P-k\rangle$ with momentum $P-k$ and mass $M_X$.

Within the twist-3 formalism, the tree-level spectator model expressions for the gluon-gluon correlator take the form:
\begin{align}
&\Phi^{i-}\left(x,\bm{k}_T,S\right)\notag\\
=&\frac{1}{x} \frac{1}{(2\pi)^3} \frac{1}{2(1-x)P^+} \mathrm{Tr} \bigg[(\slashed{P}+M)\frac{1+\gamma^5 \slashed{S}}{2}\notag\\
&\times G_{ab^\prime}^{+i\nu*}\left(k,k\right) \mathcal{Y}_{\nu,b^\prime c^\prime}^{*} G_{ab}^{+-\mu}\left(k,k\right) \mathcal{Y}_{\mu,b^\prime c} (\slashed{P}-\slashed{k}+M_X)_{cc^\prime} \bigg]\,,\label{eq:Re_Phi1}\\
&\Phi^{ij;l}\left(x,\bm{k}_T,S\right)\notag\\
=&\frac{1}{x} \frac{1}{(2\pi)^3} \frac{1}{2(1-x)P^+} \mathrm{Tr} \bigg[(\slashed{P}+M)\frac{1+\gamma^5 \slashed{S}}{2}\notag\\
&\times G_{ab^\prime}^{ij\nu*}\left(k,k\right) \mathcal{Y}_{\nu,b^\prime c^\prime}^{*} G_{ab}^{l+\mu}\left(k,k\right) \mathcal{Y}_{\mu,b^\prime c} (\slashed{P}-\slashed{k}+M_X)_{cc^\prime} \bigg]\,,\label{eq:Re_Phi2}
\end{align}
where $a,b,c,...$ are color indices and the term
\begin{align}
G_{ab}^{\mu \nu \rho}\left(p,k\right)=-\frac{i}{k^2}\left(p^\mu g^{\nu \rho}-k^\nu g^{\mu \rho}\right)\delta_{ab}\,,
\end{align}
is a specific Feynman rule for the gluon field strength tensor in the definition of the correlator~\cite{Goeke:2006ef,Collins:2011zzd}. 

The nucleon-gluon-spectator vertex $\mathcal{Y}_{\mu,b^\prime c}$ is modeled as
\begin{align}
\mathcal{Y}_{\mu,b^\prime c} \left(k^2\right)=\left[g_1\left(k^2\right) \gamma_\mu +g_2\left(k^2\right) \frac{i}{2M}\sigma_{\mu \nu} k^\nu\right]\delta_{b^\prime c}\,,\label{eq:vertex0}
\end{align}
with $\sigma_{\mu \nu}=i\left[\gamma_\mu,\gamma_\nu\right]/2$. $g_{1,2}\left(k^2\right)$ denote the nucleon-gluon-spectator couplings form factors. Although the vertex could, in principle, contain more Dirac structures, the choice in Eq.~\eqref{eq:vertex0} with the two independent Dirac structures $\gamma_\mu$ and $\sigma_{\mu \nu}$ allows us to mimic the conserved electromagnetic current of a nucleon obtained from the Gordon decomposition, given that the spectator is an on-shell spin-1/2 particle. Following previous spectator model studies of twist-2 gluon TMDs~\cite{Bacchetta:2020vty,Bacchetta:2024fci}, we adopt the dipolar form factors
\begin{align}
g_{1,2}\left(k^2\right)=\kappa_{1,2} \frac{k^2}{|k^2-\Lambda_X^2|^2}=\kappa_{1,2} \frac{k^2 (1-x)^2}{(\kt +L_X^2(\Lambda_X^2))^2}\,,\label{eq:coupling}
\end{align}
where $\kappa_{1,2}$ are free coupling parameters and $\Lambda_X$ is cut-off parameter, and 
\begin{align}
L_X^2(\Lambda_X^2)=xM_X^2+(1-x)\Lambda_X^2-x(1-x)M^2\,.
\end{align}

Since the spectator particle is on-shell, its momentum satisfies $(P-k)^2 = M_X^2$. The gluon, however, is off-shell with virtuality $k^2$. Using the on-shell condition for the spectator, $k^2$ can be expressed in terms of $x$, $\bm{k}_T$, and $M_X$ as
\begin{align}
k^2=-\frac{\kt +L_X^2(0)}{1-x}\,.\label{eq:k2}
\end{align}

The real parts of the twist-3 gluon TMDs are then obtained by applying suitably projection operators to $\Phi^{i-}$ and $\Phi^{ij;l}$ in Eqs.~\eqref{eq:Re_Phi1} and \eqref{eq:Re_Phi2}
\begin{align}
&G_3^{\perp \Re}\left(x,\kt \right)\notag\\
=&\frac{-2}{\kt } k_T^i \Phi_U^{i-}\left(x,\bm{k}_T\right)\notag\\
=&\bigg[4g_1^2 M^2 (x-1) \Big(\kt  (x-2)+(M(x-1)+M_X)\notag\\
&\times\big(M(x^2+x-2)+M_X(x-2)\big)\Big)+4g_1 g_2 xM\notag\\
&\times(1-x)(M+M_X)\big(\kt +(M(x-1)+M_X)^2\big)\notag\\
&+g_2^2\Big(2\ktquad +\kt \big(M^2(x^2+x-2)+2MM_X (x-1)x\notag\\
&+M_X^2(x^2+x+2)\big)\Big)+x\Big(M^4(1-x)^3\notag\\
&+2M^3M_X (1-x)^2 x+M^2 M_X^2 (x-1)(x^2+2)\notag\\
&+2M M_X^3(x-1)x+M_X^4(x+1)\Big)\bigg]\notag\\
&\times\bigg[32\pi^3 M^2 (x-1)x (\kt  + L_X^2(0))^2\bigg]^{-1}\,,\\
&\Delta G_{3L}^{\perp \Re}\left(x,\kt \right)\notag\\
=&\frac{-2}{iS_L \kt } \epsilon_T^{k_T i} \Phi_L^{i-}\left(x,\bm{k}_T\right)\notag\\
=&\Big(2g_1 M+g_2 (M-M_X)\Big)\Big(2g_1 M -g_2 (M+M_X)\Big)\notag\\
&\times \Big(M^2 (1-x)^2-\kt  -M_X^2\Big)\notag\\
&\times\bigg[32\pi^3 M^2  (\kt  + L_X^2(0))^2\bigg]^{-1}\,,\\
&\Delta G_{3T}^{\Re} \left(x,\kt \right)\notag\\
=&\frac{2}{iM \epsilon_T^{S_T k_T}} \bigg(k_T^i -\frac{\kt }{2 \bm{k}_T \cdot \bm{S}_T} S_T^i\bigg) \Phi_T^{i-}\left(x,\bm{k}_T\right)\notag\\
=&\bigg[(2g_1 M-g_2 (M+M_X))\Big(2g_1 M(x-1)(xM(\kt -2M^2\notag\\
&+2M_X^2)+\kt  M_X+x^2 M^2 (M+M_X)\notag\\
&+(M-M_X)^2 (M+M_X) )-g_2 \Big(\ktquad  +\kt  \big(M^2 x (x-1)\notag\\
&+M M_X (x^2-1) +M_X^2 (x+1) \big)\notag\\
&+x(M(x-1)+M_X)^2 (M^2 (x-1)+M_X^2) \Big)\Big)\bigg]\notag\\
&\times\bigg[32\pi^3 M^3 (x-1)  (\kt  + L_X^2(0))^2\bigg]^{-1}\,,\\
&\Delta G_{3T}^{\perp \Re} \left(x,\kt \right)\notag\\
=&\frac{2M}{i \bm{k}_T \cdot \bm{S}_T \epsilon_T^{k_T S_T}} S_T^i \Phi_T^{i-}\left(x,\bm{k}_T\right)\notag\\
=&(x-1)\Big(4g_1^2 M^2-4g_1 g_2 M M_X+g_2^2 (M_X^2-M^2)\Big)\notag\\
&\times \bigg[16\pi^3 (\kt  + L_X^2(0))^2\bigg]^{-1}\,,
\end{align}
and
\begin{align}
&H_3^{\perp \Re}\left(x,\kt \right)\notag\\
=&\frac{-2}{\kt } g_T^{jl} k_T^i \Phi_U^{ij;l}\left(x,\bm{k}_T\right)\notag\\
=&\Big(g_2(M+M_X)-2g_1 M\Big)^2 \Big(\kt +(M(x-1)+M_X)^2\Big)\notag\\
&\times \bigg[32\pi^3 M^2 (\kt  + L_X^2(0))^2\bigg]^{-1}\,,\\
&\Delta H_{3L}^{\perp \Re}\left(x,\kt \right)\notag\\
=&\frac{-1}{i S_L \kt } \epsilon_T^{ij}k_T^l \Phi_L^{ij;l}\left(x,\bm{k}_T\right)\notag\\
=&\bigg[\big(2g_1 M-g_2 (M+M_X)\big)\Big(-2g_1 \kt  M (x-2)\notag\\
& +2g_1 Mx(M(x-1)+M_X)^2+g_2 \kt \big(M(2-3x)\notag\\
&+M_X(x-2)\big)-g_2 x(M+M_X)(M(x-1)+M_X)^2  \Big)\bigg]\notag\\
&\times \bigg[32\pi^3 M^2 x (\kt  + L_X^2(0))^2\bigg]^{-1}\,,\\
&\Delta H_{3T}^{\Re} \left(x,\kt \right)\notag\\
=&\frac{-1}{iM \bm{k}_T \cdot \bm{S}_T} \epsilon_T^{ij}\bigg[\frac{\kt  \bm{k}_T \cdot \bm{S}_T}{2(\kt  \bm{S}_T^2-(\bm{k}_T \cdot \bm{S}_T)^2)} S_T^l\notag\\
&+\frac{\kt  \bm{S}_T^2-2(\bm{k}_T \cdot \bm{S}_T)^2}{2(\kt  \bm{S}_T^2-(\bm{k}_T \cdot \bm{S}_T)^2)}k_T^l \bigg] \Phi_T^{ij;l}(x,\bm{k}_T) \notag\\
=& \Big(2g_1 M-g_2 (M+M_X)\Big)\Big(2g_1 M (x-1)(M(x-1)+M_X)\notag\\
&-g_2 (\kt +M_X x (M(x-1)+M_X))\Big)\kt \notag\\
&\times \bigg[32\pi^3 M^3 x (\kt  + L_X^2(0))^2\bigg]^{-1}\,,\label{eq:Re_HT}\\
&\Delta H_{3T}^{\perp\Re} \left(x,\kt \right)\notag\\
=&\frac{2M}{i \bm{k}_T \cdot \bm{S}_T} \epsilon_T^{ij}\bigg[\frac{\bm{k}_T \cdot \bm{S}_T}{2(\kt  \bm{S}_T^2-(\bm{k}_T \cdot \bm{S}_T)^2)} S_T^l \notag\\
& - \frac{\bm{S}_T^2 }{2(\kt  \bm{S}_T^2-(\bm{k}_T \cdot \bm{S}_T)^2)} k_T^l \bigg]\Phi_T^{ij;l}(x,\bm{k}_T) \notag\\
=& \Big(2g_1 M-g_2 (M+M_X)\Big)\Big(2g_1 M (x-1)(M(x-1)+M_X)\notag\\
&-g_2 (\kt +M_X x (M(x-1)+M_X))\Big)\notag\\
&\times \bigg[16\pi^3 M x (\kt  + L_X^2(0))^2\bigg]^{-1}\,.\label{eq:Re_HTP}
\end{align}

We emphasize that, from Eqs.~\eqref{eq:Re_HT} and~\eqref{eq:Re_HTP}, one obtains a relation between $\Delta H_{3T}^{\Re} \left(x,\kt \right)$ and $\Delta H_{3T}^{\perp\Re} \left(x,\kt \right)$:
\begin{align}
\Delta H_{3T}^{\Re} \left(x,\kt \right)=\frac{\kt }{2M^2} \Delta H_{3T}^{\perp\Re} \left(x,\kt \right)\,. \label{eq:relation}
\end{align}
This relation is model-dependent and holds only within the specific model considered here.

\subsection{The imaginary parts of twist-3 TMDs}\label{subsection:imaginary}

\begin{figure}[htbp]
    \centering
    \includegraphics[width=0.8\columnwidth]{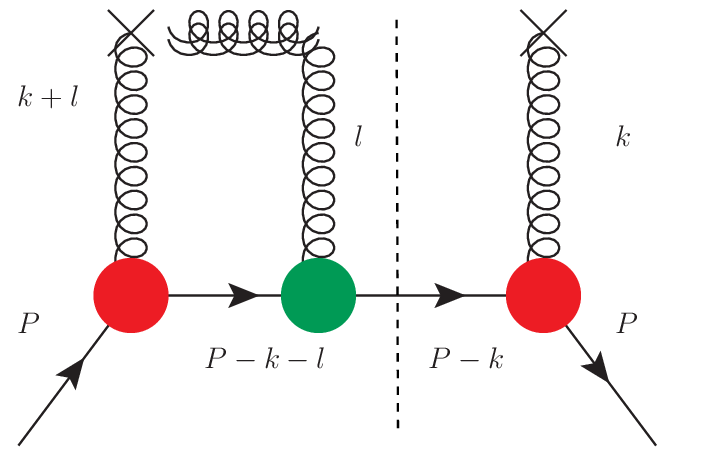}
    \caption{One-loop cut diagram for the calculation of the imaginary parts of twist-3 gluon TMDs, including a single-gluon exchange. The red blob represents the nucleon-gluon-spectator vertex, the green blob stands for the spectator-gluon-spectator vertex. Gluon lines with crosses correspond to the specific Feynman rules for the gluon field strength tensor. The eikonal propagator arising from the Wilson line is indicated by a gluon double line.}\label{fig:one-loop}
\end{figure}

In this section, we compute the imaginary parts of the twist-3 gluon TMDs by evaluating the gluon-gluon correlator corresponding to the $[+,+]$ gauge link with a future-pointing closed Wilson path.

The correlators are given by (see Fig.~\ref{fig:one-loop})
\begin{align}
&\Phi^{i-[+,+]}\left(x,\bm{k}_T,S\right)\notag\\
=&\frac{1}{x}\frac{1}{(2\pi)^3} \frac{1}{2(1-x)P^+} \mathrm{Tr}\Bigg[(\slashed{P}+M) \frac{1+\gamma^5 \slashed{S}}{2}\notag\\
&G^{+i\sigma*}(k,k) \mathcal{Y}_\sigma^{ab*}(k^2) (\slashed{P}-\slashed{k}+M_X) (g_s n_-^\alpha f^{dac})\notag\\
&\times \int \frac{d^4 l}{(2\pi)^4} \bigg(\frac{-i \mathcal{X}_\alpha^{dbe}(l^2)}{l^2-m_g^2}\bigg) \bigg(\frac{-i}{l^+ + i\epsilon}\bigg)\notag\\
&\times \frac{i(\slashed{P}-\slashed{k}-\slashed{l}+M_X)}{(P-k-l)^2-M_X^2+i\epsilon} \mathcal{Y}_\rho^{ec}((k+l)^2) G^{+-\rho}(k,k+l) \Bigg]\,,\\
&\Phi^{ij;l[+,+]}\left(x,\bm{k}_T,S\right)\notag\\
=&\frac{1}{x}\frac{1}{(2\pi)^3} \frac{1}{2(1-x)P^+} \mathrm{Tr}\Bigg[(\slashed{P}+M) \frac{1+\gamma^5 \slashed{S}}{2}\notag\\
&G^{ij\sigma*}(k,k) \mathcal{Y}_\sigma^{ab*}(k^2) (\slashed{P}-\slashed{k}+M_X) (g_s n_-^\alpha f^{dac})\notag\\
&\times \int \frac{d^4 l}{(2\pi)^4} \bigg(\frac{-i \mathcal{X}_\alpha^{dbe}(l^2)}{l^2-m_g^2}\bigg) \bigg(\frac{-i}{l^+ + i\epsilon}\bigg)\notag\\
&\times \frac{i(\slashed{P}-\slashed{k}-\slashed{l}+M_X)}{(P-k-l)^2-M_X^2+i\epsilon} \mathcal{Y}_\rho^{ec}((k+l)^2) G^{l+\rho}(k,k+l) \Bigg]\,,
\end{align}
where $\mathcal{X}_\alpha^{bde}$ is the spectator-gluon-spectator vertex, defined as
\begin{align}
\mathcal{X}_\alpha^{bde}(k^2) =f^{bde} \mathcal{Y}_\alpha^f (k^2)-i d^{bde} \mathcal{Y}_\alpha^d (k^2)\,.
\end{align}
The vertices $\mathcal{Y}_\alpha^{f,d} (k^2)$ are not required to coincide with the nucleon-gluon-spectator vertex $\mathcal{Y}_\alpha (k^2)$ defined in Eq.~\eqref{eq:vertex0}. For simplicity, however, we assume $\mathcal{Y}_\alpha^{f,d} (k^2)=\mathcal{Y}_\alpha (k^2)$. Under this assumption, the $f$-type and $d$-type gluon TMDs are related by a constant color factor:
\begin{align}
[+,+]/[+,-]=(f^{acd}f^{dca})/((-id^{acd})(-id^{dca}))=9/5\,.
\end{align}
Meanwhile, the cases $[+,+]$ ($[-,-]$) and $[+,-]$ ($[-,+]$) are related by the modified-universality relations given in Eqs.~\eqref{eq:gauge1} and~\eqref{eq:gauge2}. We therefore present our results exclusively for the $f$-type gluon TMDs.

To extract the imaginary parts for twist-3 gluon TMDs, we consider the cuts through the eikonal line and the spectator line inside the loop diagram. Applying the Cutkosky rules, we can make the following replacement
\begin{align}
\frac{1}{l^++i\epsilon}\to&-2\pi i \delta(l^+)\,,\\
\frac{1}{(P-k-l)^2-M_X^2+i\epsilon}\to & -2\pi i \delta((P-k-l)^2-M_X^2)\,.
\end{align}
Furthermore, we make use of the spectator model relation Eq.~\eqref{eq:k2}, which can be applied to the momenta $k^2$, $l^2$, or $(k+l)^2$, appearing in the propagators and form factors.

The general structure for the imaginary part of a twist-3 gluon TMD, denoted by $F^{\Im} \left(x,\kt \right)$, can then be organized as follows:
\begin{align}
F^{\Im} \left(x,\kt \right)=&\mathrm{Im}\frac{(1-x)^4 P^+}{(2\pi)^3 [\kt +L_X^2(\Lambda_X)]^2}\sum_{i,j,k}^{1,2} \sum_{l=1}^{8}\notag\\
&\times \mathcal{C}_{ijk}^{[F],l}\left(x,\kt \right)~\mathcal{D}_l\left(x,\kt \right) g_s \kappa_i \kappa_j \kappa_k\,,\label{eq:F_IM}
\end{align}
where $\kappa_{i,j,k}$ are the coupling parameters from the dipolar form factors given in Eq.~\eqref{eq:coupling}. The symbol $\mathcal{D}_l\left(x,\kt\right)$ represent eight different master integrals listed in Appendix~\ref{appendix1}, and $\mathcal{C}_{ijk}^{[F],l}\left(x,\kt\right)$ are the corresponding coefficients provided in Appendix~\ref{appendix2}. It is noteworthy that the non-vanishing imaginary parts of the twist-3 TMDs arise directly from the fact that the diagram in Fig.~\ref{fig:one-loop} and its complex conjugate yield distinct imaginary contributions, rather than simply contributions of opposite sign.

\subsection{Dynamical twist-3 TMDs}
\begin{figure}[htbp]
    \centering
    \includegraphics[width=0.8\columnwidth]{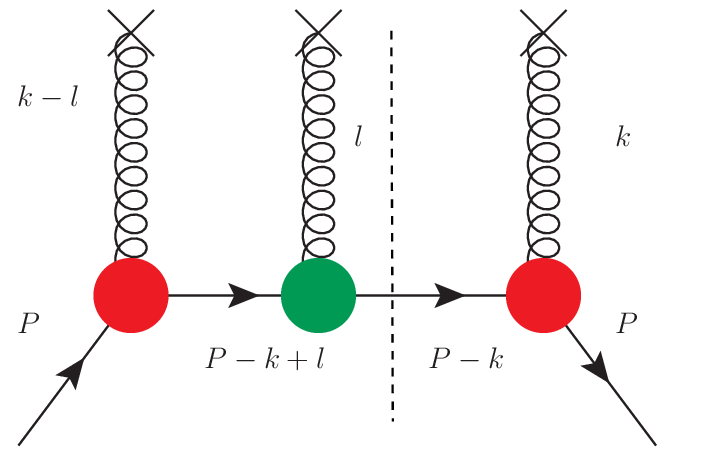}
    \caption{Cut diagram for the calculation of the three-gluon correlator in the spectator model. The red blob represents the nucleon-gluon-spectator vertex. The green blob stands for the spectator-gluon-spectator vertex. Gluon lines with crosses correspond to specific Feynman rules for the gluon field tensor. }\label{fig:tri-gluon}     
\end{figure}

The dynamical collinear twist-3 gluon distributions are defined as the light-cone correlation functions of three field strength tensors ($f$-type distribution)~\cite{Beppu:2010qn,Ji:1992eu,Koike:2019zxc}
\begin{align}
&N_f^{\alpha \beta \gamma}\left(x_1,x_2\right)\notag\\
=&i \int \frac{d\lambda}{2\pi} \frac{d\mu}{2\pi} e^{i\lambda x_1} e^{i \mu (x_2-x_1)}\notag\\
&\times \langle PS|if^{acb} F_a^{+\alpha}(0) g_s F_c^{+\gamma}(\mu n_-) F_b^{+\beta}(\lambda n_-)|PS \rangle\notag\\
=&2iM \Big[-g_T^{\alpha \beta} \epsilon_T^{\gamma S_T}N(x_1,x_2) +g_T^{\alpha \gamma} \epsilon_T^{\beta S_T} N(x_2,x_2-x_1)\notag\\
&+g_T^{\beta \gamma} \epsilon_T^{\alpha S_T} N(x_1, x_1-x_2) \Big]\,.
\end{align}
The function $N(x_1,x_2)$ satisfies the symmetry relations $N(x_1,x_2)=N(x_2,x_1)$ and $N(-x_1,-x_2)=-N(x_1,x_2)$, and is supported over the kinematic region $|x_{1,2}|<1$ and $|x_1-x_2|<1$.

These dynamical distributions are connected to the intrinsic integrated twist-3 TMDs in Eqs.~\eqref{eq:phi1}-\eqref{eq:phi2} through the equation of motion relation~\cite{Koike:2019zxc}
\begin{align}
\frac{1}{2} x^2& \Delta H_{3T}(x)=\frac{1}{2}\Delta G_T^{(1)}\left(x\right)\notag\\
&+2\int dx_1 \mathcal{P}\frac{1}{x-x_1}\{N(x_1,x)+N(x,x-x_1)\}\,. \label{eq:motion}
\end{align}
The integrated twist-3 TMD $\Delta H_{3T}(x)$ and the kinematical function $\Delta G_T^{(1)}\left(x\right)$ are given by
\begin{align}
\Delta H_{3T}(x)=&\int d^2 \bm{k}_T \Delta H_{3T}(x,\kt )\,,\\
\Delta G_T^{(1)}\left(x\right)=&\int d^2 \bm{k}_T \frac{\kt }{2M^2} \Delta G_T\left(x,\kt \right)\,,\\
=&-\int d^2 \bm{k}_T \frac{\kt }{2M^2} g_{1T}\left(x,\kt \right)\,,
\end{align}
where $\Delta G_T\left(x,\kt \right)$ (or $g_{1T}\left(x,\kt \right)$) denotes the twist-2 worm-gear gluon TMD.

The integrated dynamical gluon TMD, $\tilde{N}(x)=\int d^2 \bm{k}_T \tilde{N}(x,\kt )$, is related to the collinear twist-3 dynamical function via
\begin{align}
\tilde{N}(x)=\int dx_1 \mathcal{P}\frac{1}{x-x_1}\{N(x_1,x)+N(x,x-x_1)\}\,.
\end{align}
The dynamical function $\tilde{N}\left(x,\kt \right)$ can be extracted from the three-gluon correlator $\tilde{\Phi}^{\alpha \beta \gamma}\left(x,\bm{k}_T\right)$ by the following projection:
\begin{align}
\tilde{N}\left(x,\kt \right)=\frac{k_T^\beta g_T^{\alpha \gamma}-k_T^\gamma g_T^{\alpha \beta}}{2M \epsilon_T^{k_T S_T}} \tilde{\Phi}^{\alpha \beta \gamma}\left(x,\bm{k}_T\right)\,.
\end{align}

The three-gluon correlator $\tilde{\Phi}^{\alpha \beta \gamma}\left(x,\bm{k}_T\right)$ can be written as~\cite{Lu:2015wja,Yang:2016mxl}
\begin{align}
&\tilde{\Phi}^{\alpha \beta \gamma}\left(x,\bm{k}_T\right)\notag\\
=&\frac{1}{x P^+}\frac{1}{(2\pi)^3} \frac{1}{2(1-x)P^+} \mathrm{Tr}\Bigg[(\slashed{P}+M) \frac{1+\gamma^5 \slashed{S}}{2}\notag\\
&G^{+ \alpha \sigma*}(k,k) \mathcal{Y}_\sigma^{ab*}(k^2) (\slashed{P}-\slashed{k}+M_X) (i g_s  f^{dac})\notag\\
&\times \int \frac{d^4 l}{(2\pi)^4} \bigg(\frac{ G^{+ \gamma \mu }(l,l) \mathcal{Y}_\mu(l^2) f^{dbe}}{l^2-m_g^2}\bigg) \bigg(\frac{-i}{l^+ + i\epsilon}\bigg)\notag\\
&\times \frac{i(\slashed{P}-\slashed{k}+\slashed{l}+M_X)}{(P-k+l)^2-M_X^2+i\epsilon} \mathcal{Y}_\rho^{ec}((k-l)^2) G^{+\beta\rho}(k,k-l) \Bigg]\,.
\end{align}
Similarly to the calculation of the imaginary parts of twist-3 TMDs in Sec.~\ref{subsection:imaginary}, the dynamical gluon TMD $\tilde{N}\left(x,\kt \right)$ is expressed as:
\begin{align}
\tilde{N}\left(x,\kt \right)=&\frac{(1-x)^4 P^+}{(2\pi)^3 [\kt +L_X^2(\Lambda_X)]^2}\sum_{i,j,k}^{1,2} \sum_{l=1}^{11}\notag\\
&\times \mathcal{C}_{ijk}^{[\tilde{N}],l}\left(x,\kt \right)~\mathcal{D}_l\left(x,\kt \right) g_s \kappa_i \kappa_j \kappa_k\,,\label{eq:N_dy}
\end{align}
where the coefficients $\mathcal{C}_{ijk}^{[\tilde{N}],l}\left(x,\kt \right)$ for $l=1,2,...,11$ are listed in Appendix~\ref{appendix2}.

Finally, in our model, the spectator mass $M_X$ can take a continuous range of real values, described by the spectral function~\cite{Bacchetta:2020vty}:
\begin{align}
\rho\left(M_X\right)=\mu^{2a}\left[\frac{A}{B+\mu^{2b}}+\frac{C}{\pi \sigma} e^{-\frac{(M_X-D)^2}{\sigma^2}}\right]\,,
\end{align}
where $\mu^2=M_X^2-M^2$ and $\{X\}\equiv\{A,B,a,b,C,D,\sigma\}$ are free model parameters. This spectral function models the spectator mass as a resonance peak superimposed on a smooth background. The gluon TMDs are then weighted by the spectral function $\rho\left(M_X\right)$:
\begin{align}
F\left(x,\kt \right)=\int_{M}^{\infty}dM_X\ \rho\left(M_X\right) F\left(x,\kt ;M_X\right)\,,
\end{align}
which provides an effective way in a spectator model to account for $q\bar{q}$ contributions to spectator configurations that become energetically available at large $M_X$.

\section{Numerical results}\label{section3}

\begin{figure*}[htbp]
    \centering
    \begin{tabular}{c@{\hspace{1em}}c}
    \includegraphics[width=1\columnwidth]{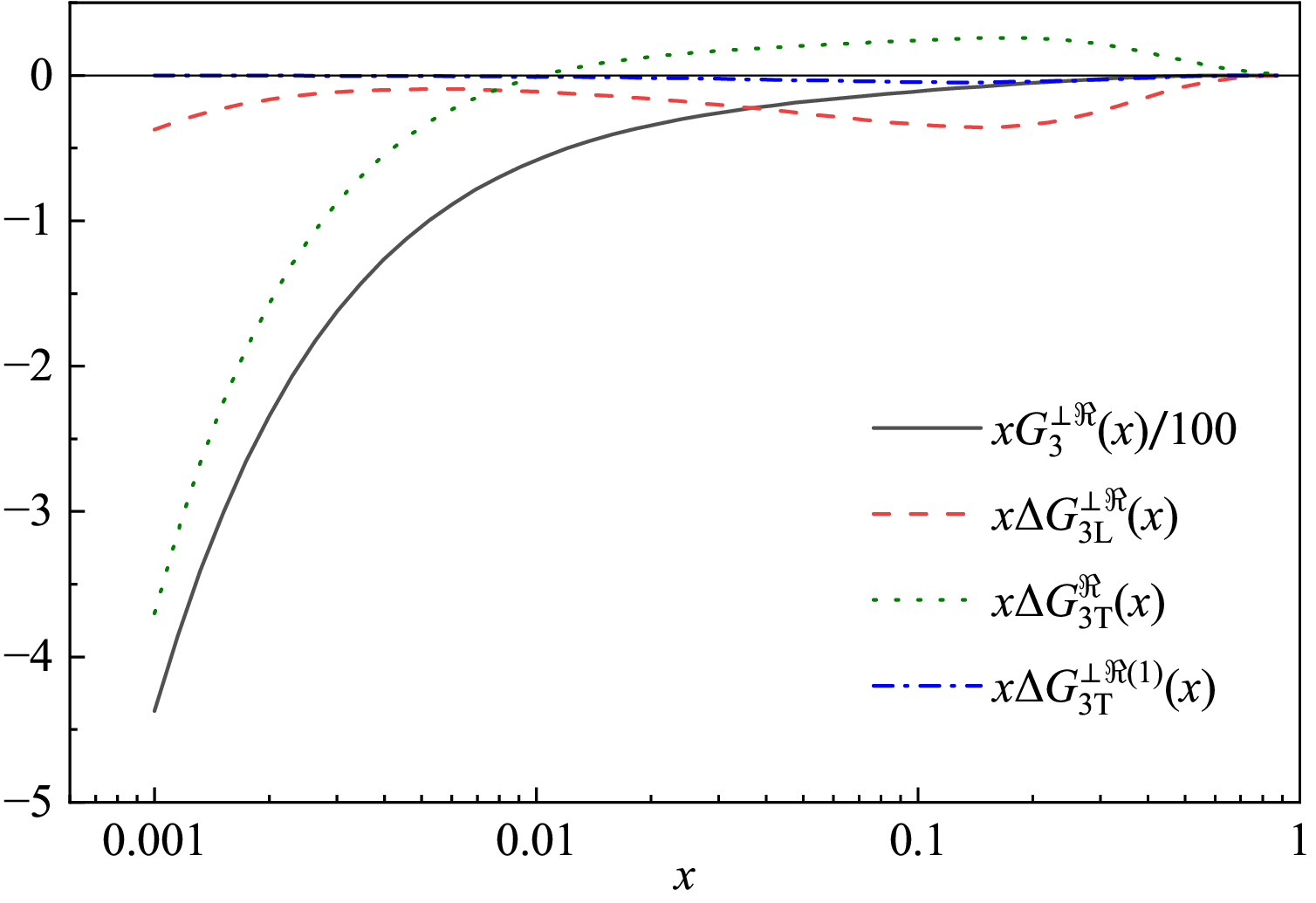}&\includegraphics[width=1\columnwidth]{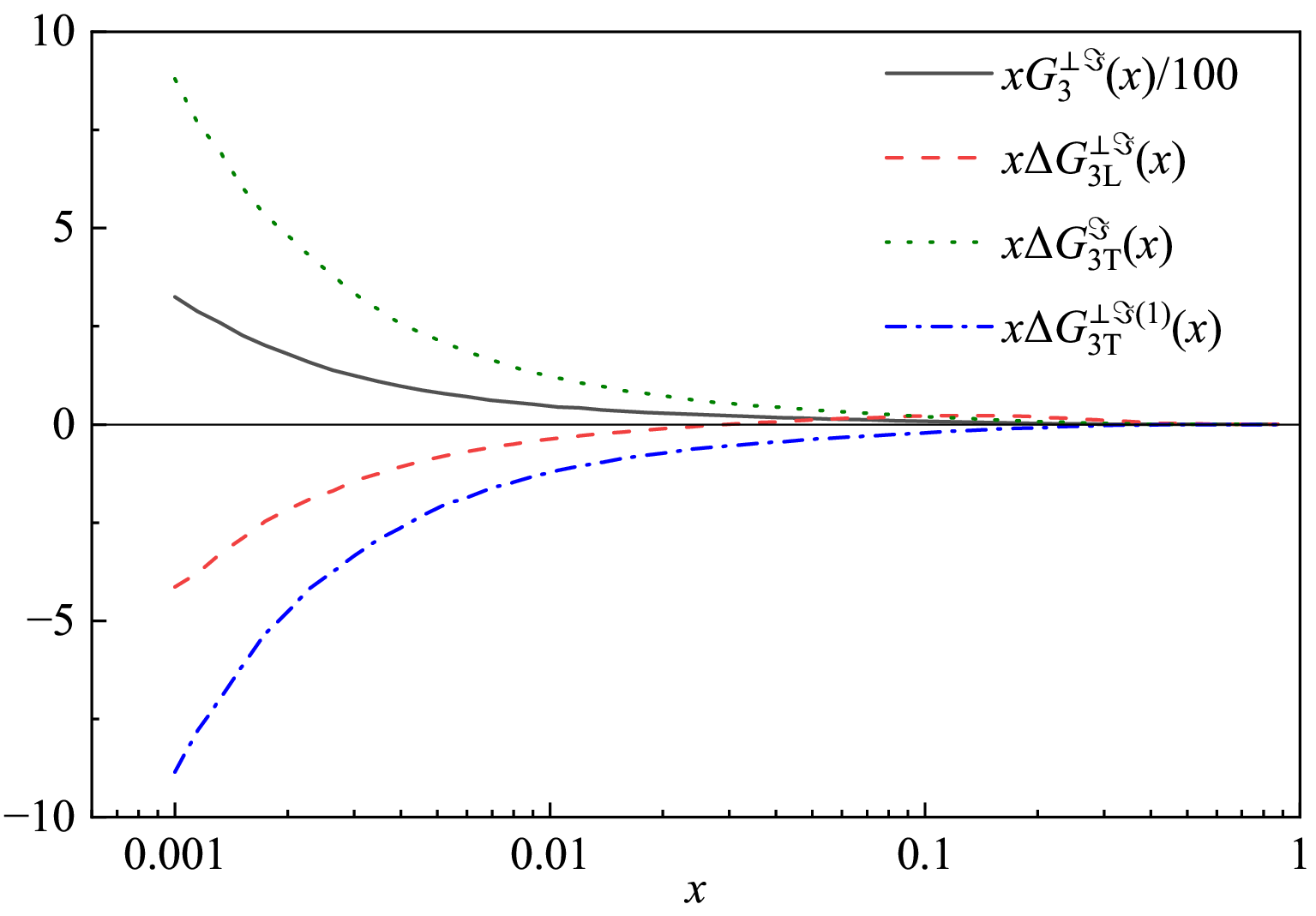}\\
    \includegraphics[width=1\columnwidth]{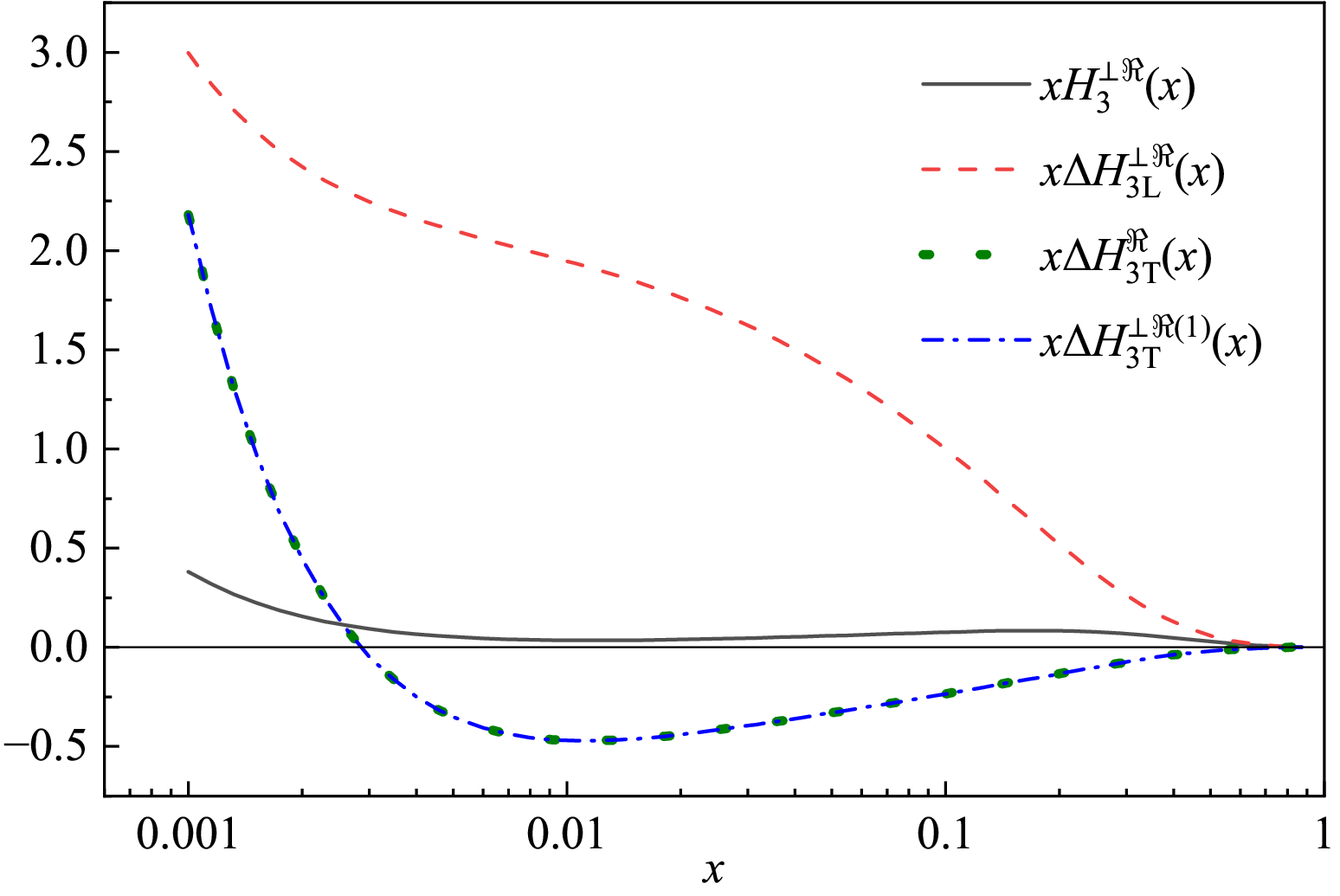}&\includegraphics[width=1\columnwidth]{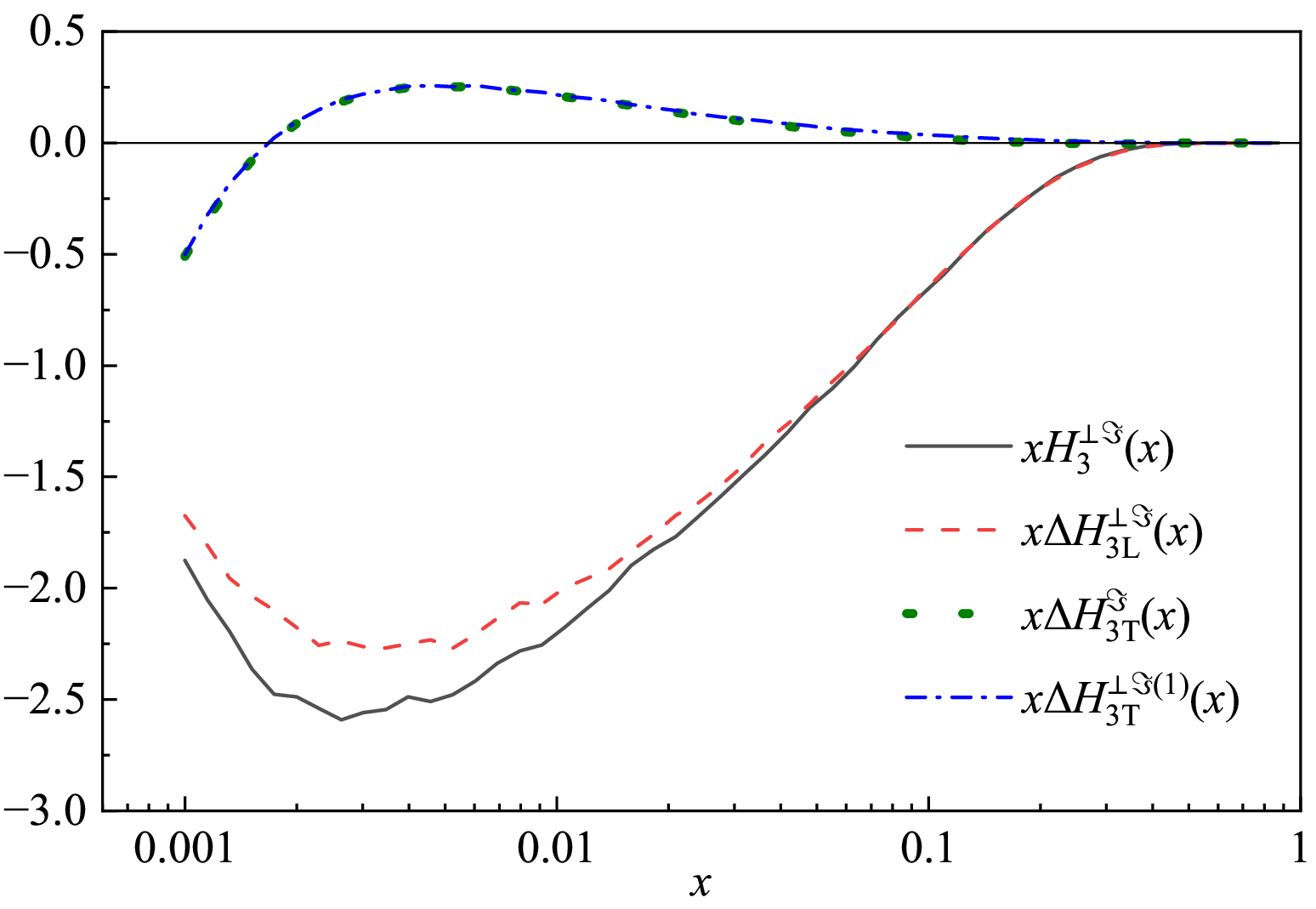}
    \end{tabular}
    \caption{Upper-left: Real parts of four integrated twist-3 gluon TMDs obtained from $\Phi^{i-}$ as functions of $x$. Upper-right:  Corresponding imaginary parts. Lower-left: Real parts of four integrated twist-3 gluon TMDs obtained from $\Phi^{ij;l}$ as functions of $x$. Lower-right:  Corresponding imaginary parts.}\label{fig:x}     
\end{figure*}

\begin{figure*}[htbp]
    \centering
    \includegraphics[width=2\columnwidth]{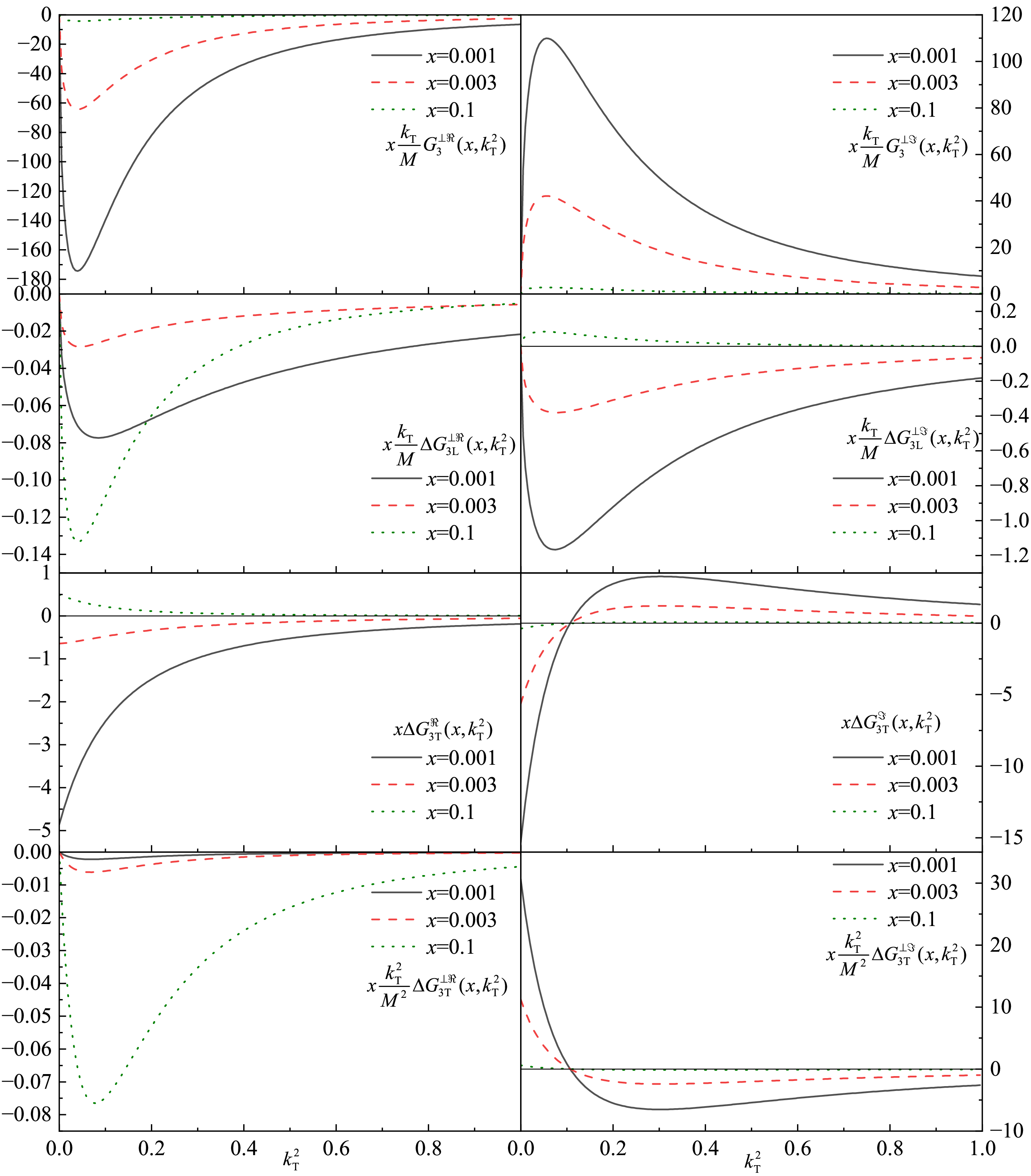}
    \caption{Twist-3 gluon TMDs obtained from $\Phi^{i-}$ as functions of $\kt$ for $x=0.001$, $x=0.003$, $x=0.1$. From top to bottom, the panels show $x\bm{k}_T/M~G_3^{\perp}$, $x\bm{k}_T/M~\Delta G_{3L}^{\perp}$, $x\Delta G_{3T}$, and $x\bm{k}_T^2/M^2~\Delta G_{3T}^{\perp}$. Left panels display the real parts, and right panels show the imaginary parts.} \label{fig:kt_G}   
\end{figure*}

\begin{figure*}[htbp]
    \centering
    \includegraphics[width=2\columnwidth]{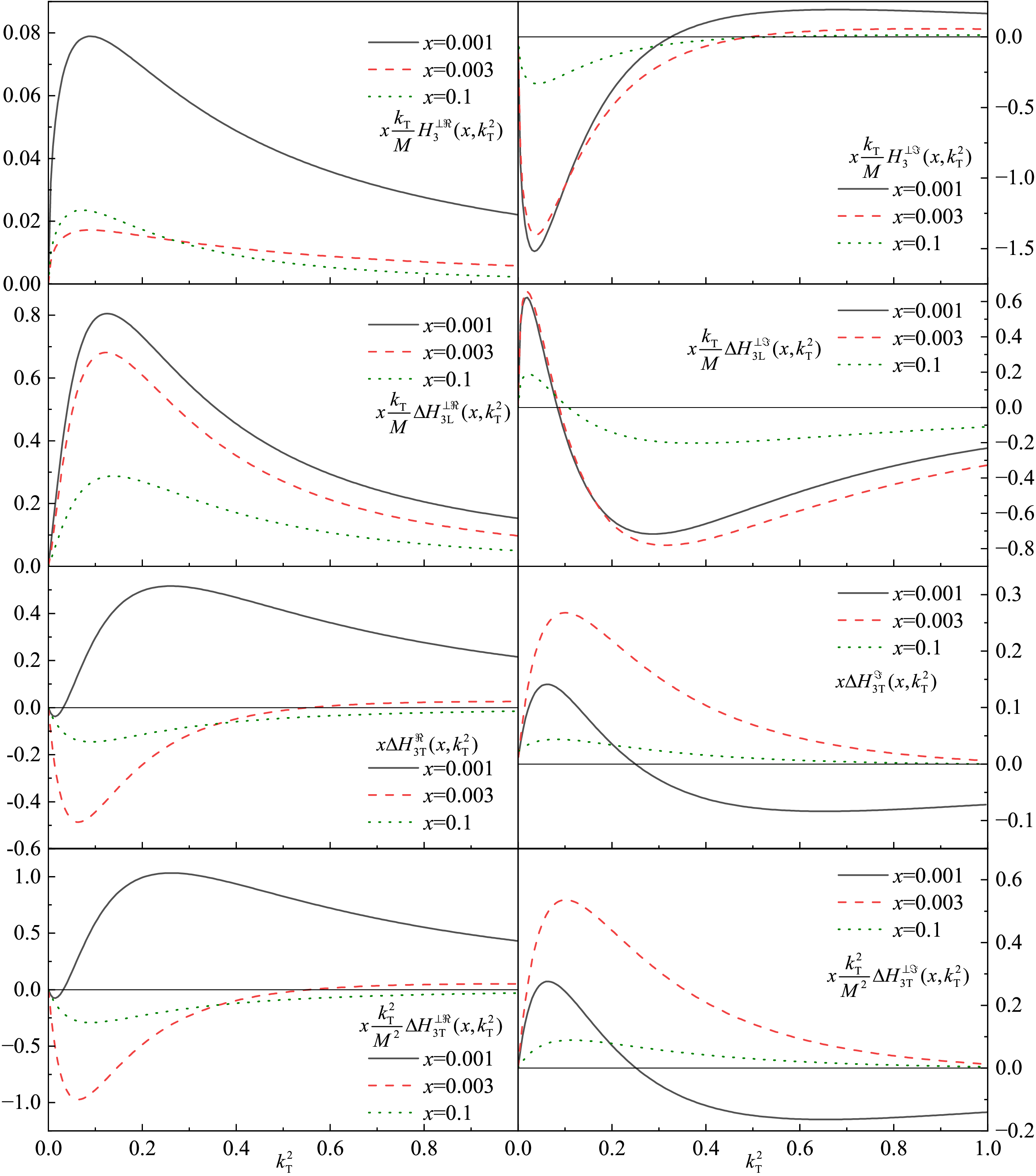}
    \caption{Twist-3 gluon TMDs obtained from $\Phi^{ij;l}$ as functions of $\kt$ for $x=0.001$, $x=0.003$, $x=0.1$. From top to bottom, the panels show $x\bm{k}_T/M~H_3^{\perp}$, $x\bm{k}_T/M~\Delta H_{3L}^{\perp}$, $x\Delta H_{3T}$, and $x\bm{k}_T^2/M^2~\Delta H_{3T}^{\perp}$. Left panels show the real parts, and right panels show the imaginary parts.}\label{fig:kt_H}
\end{figure*}

The model parameters are fixed to established numerical values taken from the literature. As detailed in Ref.~\cite{Bacchetta:2020vty}, these parameters were determined by simultaneously fitting the NNPDF3.1sx parametrization for the unpolarized $xf_1^g$ and the NNPDFpol1.1 parametrization for the helicity $xg_1^g$ at $Q_0 = 1.64\,\mathrm{GeV}$. The 68\% uncertainties were obtained by excluding the largest and smallest 16\% of all 100 replica values, which corresponds to $1\sigma$ standard deviation for a Gaussian distribution. Parameter $B$ is fixed to $B=2.1$ in all numerical calculations, as specified in Table.~\ref{table:parm}. The column labeled "replica 11" lists the parameters of the most representative replica. 
In this section, we present numerical results for the twist-3 gluon TMDs using the parameter set of replica 11. For all calculations, the strong coupling constant is fixed at $g_s=\sqrt{\alpha_s(Q_0)}=0.576$, consistent with the input scale $Q_0$.
\begin{table}[H]
\centering
\caption{Mean values and uncertainties of the fitted model parameters (central column) and the corresponding values for replica 11 (rightmost column). Parameter $B$ is fixed to $B=2.1$}\label{table:parm}
    \setlength{\tabcolsep}{0.4cm}{
    \begin{tabular}{ccc}
    \hline
    Parameter & Mean & Replica 11 \\
    \hline
    $\kappa_1$ &1.51$\pm$0.16 &1.46 \\
    $\kappa_2$ &0.414$\pm$0.036 &0.414 \\
    $\Lambda_X$&0.472$\pm$0.058 &0.448 \\
    $a$&0.82$\pm$0.21 &0.78 \\
    $b$&1.43$\pm$0.23 &1.38 \\
    $A$&6.1$\pm$2.3 &6.0 \\
    $C$&371$\pm$58 &346 \\
    $D$&0.548$\pm$0.081 &0.548 \\
    $\sigma$&0.52$\pm$0.14 &0.50 \\
    \hline
    \end{tabular}}
\end{table}

In Fig.~\ref{fig:x}, we present the integrated results for the full set of twist-3 gluon TMDs. The distributions $\Delta G_{3T}^{\perp}$ and $\Delta H_{3T}^{\perp}$ are shown as their first moment $F^{(1)}(x)=\int d^2 \bm{k}_T~\kt/2M^2~F(x,\kt)$. The numerical results indicate that the magnitudes of these twist-3 TMDs are comparable and can be sizable. Owing to the relation given in Eq.~\eqref{eq:relation}, the analytical expressions for $x\Delta H_{3T}^\Re$ and $x\Delta H_{3T}^{\perp \Re (1)}$ are equal. We also observe that $x\Delta H_{3T}^\Im$ and $x\Delta H_{3T}^{\perp \Im (1)}$ are approximately equal, whereas $x\Delta G_{3T}^\Im$ and $x\Delta G_{3T}^{\perp \Im (1)}$ are nearly opposite in sign. Furthermore, in the region $0.001<x<0.01$, both the real and imaginary parts of $x\Delta H_{3T}$ and $x\Delta H_{3T}^{\perp  (1)}$ exhibit a node and change sign.

Figure~\ref{fig:kt_G} displays the twist-3 gluon TMDs derived from the correlator $\Phi^{i-}$ as functions of $\bm{k}_T^2$ at $x = 0.001$ (solid), $x = 0.003$ (dashed), and $x = 0.1$ (dotted). All functions exhibit a pronounced falloff as $\bm{k}_T^2$ increases. $x\bm{k}_T/M\, G_3^{\perp\Re}$ and $x\bm{k}_T/M\, G_3^{\perp\Im}$ show opposite trends, and each displays a peak for $\bm{k}_T^2 < 0.2\,\mathrm{GeV}^2$; these peaks shift toward smaller $\bm{k}_T^2$ as $x$ increases. Similarly, a node appears around $\bm{k}_T^2 \approx 0.1\,\mathrm{GeV}^2$ for $x\Delta G_{3T}^{\Im}$ and $x\bm{k}_T^2/M^2\,\Delta G_{3T}^{\perp\Im}$, as these two distributions have opposite tendencies and change sign at that point.

In Fig.~\ref{fig:kt_H}, we show the twist-3 gluon TMDs derived from the correlator $\Phi^{ij;l}$, as functions of $\kt$ at $x=0.001$(solid), $x=0.003$(dashed), and $x=0.1$(dotted). These distributions also fall off rapidly with increasing $\kt$. While $x\bm{k}_T/M~H_3^{\perp\Re}$ and $x\bm{k}_T/M~\Delta H_{3L}^{\perp\Re}$ remain positive over the whole $\kt$ range, $x\bm{k}_T/M~H_3^{\perp\Im}$ and $x\bm{k}_T/M~\Delta H_{3L}^{\perp\Im}$ change sign as $\kt$ increases. Interestingly, at $x=0.001$, both $x\Delta H_{3T}^\Im$ and $x\kt/M^2~\Delta H_{3T}^{\perp \Im}$ change sign, exhibiting a node at $\kt\approx 0.25~\mathrm{GeV}^2$, whereas for $x=0.003$ and $x=0.1$ they are positive over the entire $\kt$ region.

\begin{figure}[htbp]
    \centering
    \includegraphics[width=1\columnwidth]{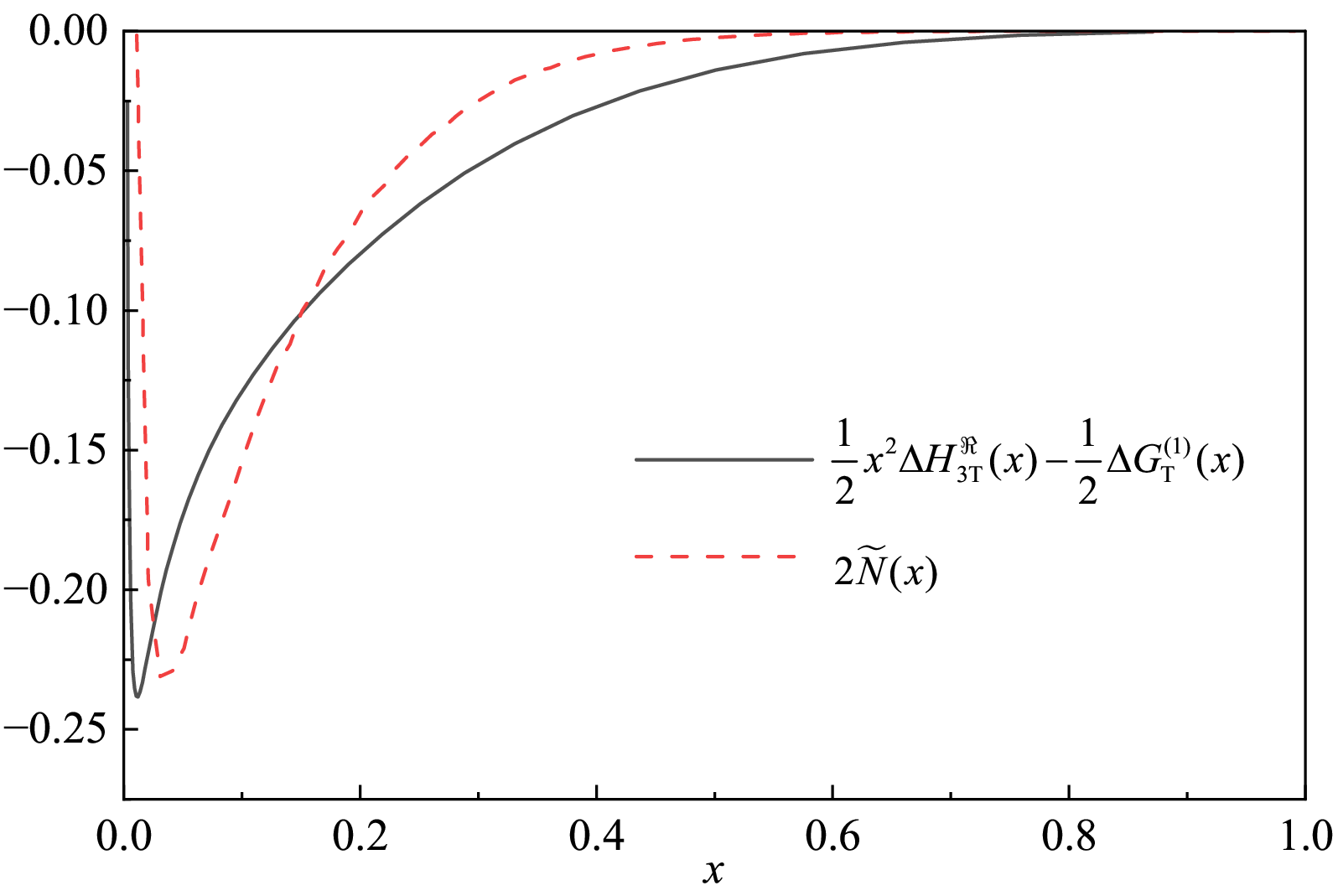}
    \caption{Dynamical twist-3 TMD $2\tilde{N}(x)$ compared with $\frac{1}{2}x^2 \Delta H_{3T}^{\Re}-\frac{1}{2}\Delta G_T^{(1)}(x)$. } \label{fig:motion}   
\end{figure}

According to Eq.~\eqref{eq:motion}, the twist-3 gluon TMDs should satisfy the  equation of motion relation, which is a model-independent result derived from QCD~\cite{Koike:2019zxc}. In a spectator model, however, a direct analytical comparison is not straightforward because $\Delta H_{3T}^\Re (x)$ from Eq.~\eqref{eq:Re_HT}, $\tilde{N}(x)$ from Eq.~\eqref{eq:N_dy}, and $\Delta G_T(x)$ from Ref.~\cite{Bacchetta:2020vty} originate from different diagrams. We therefore perform a numerical test of Eq.~\eqref{eq:motion}. In Fig.~\ref{fig:motion} we compare $\frac{1}{2}x^2 \Delta H_{3T}^{\Re} - \frac{1}{2}\Delta G_T^{(1)}(x)$ (solid curve) and $2\tilde{N}(x)$ (dashed curve). The two curves are in close agreement, indicating that the relation is approximately satisfied within the model. This agreement provides a valuable cross-check on the validity of our calculation.

\section{Conclusion}\label{section4}

In this paper, we presented a comprehensive calculation of twist-3 gluon TMDs in a nucleon within a spectator model framework. The model assumes that an on-shell nucleon can emit a time-like off-shell gluon, while the remaining system is treated as a single on-shell spectator particle. The spectator mass is treated via a spectral function, allowing it to take a continuous range of real values. 

We investigated the complete set of eight twist-3 gluon TMDs, which arise from the gluon-gluon correlators $\Phi^{i-}$ and $\Phi^{ij;l}$. These twist-3 gluon TMDs are organized such that the T-even functions correspond to the real parts and the T-odd functions correspond to the imaginary parts. We obtained the T-even parts from tree-level diagrams, while the one-loop diagrams yield the T-odd parts.

We derived complete analytical expressions and presented numerical results for all twist-3 gluon TMDs. The model parameters were determined by simultaneously fitting the NNPDF3.1sx parametrization for the unpolarized $xf_1^{g}$ and the NNPDFpol1.1 parametrization for the helicity $xg_1^g$ at $Q_0=1.64~\mathrm{GeV}$. In addition to these twist-3 gluon TMDs, we also calculated the dynamical gluon TMDs and the first moment of the twist-2 gluon TMDs. These three sets of functions are not independent but are connected by the equation of motion relation derived from QCD. Our numerical analysis shows that this relation holds approximately within our model, providing a valuable crosscheck on the validity of our calculation.

Our study may provide valuable theoretical insights into the intrinsic properties of twist-3 gluon TMDs, which are directly relevant to the interpretation of SSAs in processes such as $ep^\uparrow \to eDX$, $p^\uparrow p \to DX$, $p^\uparrow p \to \{\gamma,\gamma^*\}X$ and also $A_{LT}$ for $\vec{p} p^\uparrow \to DX$.

\section*{Acknowledgements}
This work is partially supported by the National Natural Science Foundation of China under grant number 12150013. Xiupeng Xie is also supported by the SEU Innovation Capability Enhancement Plan for Doctoral Students under grant number CXJH$\_$SEU 25138.

\begin{widetext}

\appendix

\section{Master integrals}\label{appendix1}

In the following, we list the master integrals involved in the expressions of our twist-3 gluon TMDs. We first define
\begin{align}
T_h(|\bm{k}_T|)=\mathrm{tanh}^{-1}\sqrt{\frac{\kt }{\kt +4L_X^2(\Lambda_X^2)}}\,.
\end{align}
Then, we have
\begin{align}
\mathcal{D}_1 \left(x,\kt \right)=&\frac{1}{2P^+}\int \frac{d^2 \bm{l}_T}{(2\pi)^2} \frac{1}{[\bm{l}_T^2+L_X^2(\Lambda_X^2)]^2} \frac{1}{[(\bm{l}_T+\bm{k}_T)^2+L_X^2(\Lambda_X^2)]^2}\notag\\
=&\frac{1}{8\pi P^+} \bigg[2 \frac{1-2L_X^2(\Lambda_X^2)/\kt }{L_X^2(\Lambda_X^2)[\kt +4L_X^2(\Lambda_X^2)]^2}+16 \frac{\kt +L_X^2(\Lambda_X^2)}{|\bm{k}_T|^3 [\kt +4L_X^2(\Lambda_X^2)]^{5/2}}T_h(|\bm{k}_T|)\bigg]\,,
\end{align}
\begin{align}
\mathcal{D}_2 \left(x,\kt \right)=&\frac{1}{2P^+}\int \frac{d^2 \bm{l}_T}{(2\pi)^2} \frac{\bm{l}_T \cdot \bm{k}_T}{\kt } \frac{1}{[\bm{l}_T^2+L_X^2(\Lambda_X^2)]^2} \frac{1}{[(\bm{l}_T+\bm{k}_T)^2+L_X^2(\Lambda_X^2)]^2} \equiv -\frac{1}{2} \mathcal{D}_1 \left(x,\kt \right)\,,
\end{align}
\begin{align}
\mathcal{D}_3 \left(x,\kt \right)=&\frac{1}{2P^+}\int \frac{d^2 \bm{l}_T}{(2\pi)^2} \frac{(\bm{l}_T \cdot \bm{k}_T)^2}{|\bm{k}_T|^4} \frac{1}{[\bm{l}_T^2+L_X^2(\Lambda_X^2)]^2} \frac{1}{[(\bm{l}_T+\bm{k}_T)^2+L_X^2(\Lambda_X^2)]^2}\notag\\
=&\frac{1}{8\pi P^+} \bigg[ \frac{1+3L_X^2(\Lambda_X^2)/\kt +8L_X^4(\Lambda_X^2)/\ktquad }{L_X^2(\Lambda_X^2)[\kt +4L_X^2(\Lambda_X^2)]^2}+2 \frac{\ktquad -6\kt  L_X^2(\Lambda_X^2)-16L_X^4(\Lambda_X^2)}{|\bm{k}_T|^5 [\kt +4L_X^2(\Lambda_X^2)]^{5/2}}T_h(|\bm{k}_T|)\bigg]\,,
\end{align}
\begin{align}
\mathcal{D}_4 \left(x,\kt \right)=&\frac{1}{2P^+}\int \frac{d^2 \bm{l}_T}{(2\pi)^2} \frac{\bm{l}_T^2}{\kt } \frac{1}{[\bm{l}_T^2+L_X^2(\Lambda_X^2)]^2} \frac{1}{[(\bm{l}_T+\bm{k}_T)^2+L_X^2(\Lambda_X^2)]^2}\notag\\
=&\frac{1}{8\pi P^+} \bigg[ \frac{1+2L_X^2(\Lambda_X^2)/\kt +4L_X^4(\Lambda_X^2)/\ktquad }{L_X^2(\Lambda_X^2)[\kt +4L_X^2(\Lambda_X^2)]^2}+4 \frac{\ktquad -4L_X^4(\Lambda_X^2)}{|\bm{k}_T|^5 [\kt +4L_X^2(\Lambda_X^2)]^{5/2}}T_h(|\bm{k}_T|)\bigg]\,,
\end{align}
\begin{align}
\mathcal{D}_5 \left(x,\kt \right)=&\frac{1}{2P^+}\int \frac{d^2 \bm{l}_T}{(2\pi)^2} \frac{\bm{l}_T \cdot \bm{k}_T}{\kt } \frac{\bm{l}_T^2}{\kt } \frac{1}{[\bm{l}_T^2+L_X^2(\Lambda_X^2)]^2} \frac{1}{[(\bm{l}_T+\bm{k}_T)^2+L_X^2(\Lambda_X^2)]^2}\notag\\
=&-\frac{1}{8\pi P^+} \bigg[ \frac{1+5L_X^2(\Lambda_X^2)/\kt +10L_X^4(\Lambda_X^2)/\ktquad }{L_X^2(\Lambda_X^2)[\kt +4L_X^2(\Lambda_X^2)]^2}-8 \frac{L_X^2(\Lambda_X^2)[2\kt +5L_X^2(\Lambda_X^2)]}{|\bm{k}_T|^5 [\kt +4L_X^2(\Lambda_X^2)]^{5/2}}T_h(|\bm{k}_T|)\bigg]\,,
\end{align}
\begin{align}
\mathcal{D}_6 \left(x,\kt \right)=&\frac{1}{2P^+}\int \frac{d^2 \bm{l}_T}{(2\pi)^2} \frac{|\bm{l}_T|^4}{|\bm{k}_T|^4} \frac{1}{[\bm{l}_T^2+L_X^2(\Lambda_X^2)]^2} \frac{1}{[(\bm{l}_T+\bm{k}_T)^2+L_X^2(\Lambda_X^2)]^2}\notag\\
=&\frac{1}{8\pi P^+} \bigg[ \frac{1+6L_X^2(\Lambda_X^2)/\kt +10L_X^4(\Lambda_X^2)/\ktquad -4L_X^6(\Lambda_X^2)/\bm{k}_T^6}{L_X^2(\Lambda_X^2)[\kt +4L_X^2(\Lambda_X^2)]^2}\notag\\
&-8 \frac{L_X^2(\Lambda_X^2)[\ktquad +2\kt  L_X^2(\Lambda_X^2)-2L_X^4(\Lambda_X^2)]}{|\bm{k}_T|^7 [\kt +4L_X^2(\Lambda_X^2)]^{5/2}}T_h(|\bm{k}_T|)\bigg]\,,
\end{align}
\begin{align}
\mathcal{D}_7 \left(x,\kt \right)=&\frac{1}{2P^+}\int \frac{d^2 \bm{l}_T}{(2\pi)^2} \frac{(\bm{l}_T \cdot \bm{k}_T)^2}{\ktquad } \frac{\bm{l}_T^2}{\kt } \frac{1}{[\bm{l}_T^2+L_X^2(\Lambda_X^2)]^2} \frac{1}{[(\bm{l}_T+\bm{k}_T)^2+L_X^2(\Lambda_X^2)]^2}\notag\\
=&\frac{1}{8\pi P^+} \bigg[ \frac{1+6L_X^2(\Lambda_X^2)/\kt +9L_X^4(\Lambda_X^2)/\ktquad -8L_X^6(\Lambda_X^2)/\bm{k}_T^6}{L_X^2(\Lambda_X^2)[\kt +4L_X^2(\Lambda_X^2)]^2}\notag\\
&- \frac{\bm{k}_T^6+14\ktquad  L_X^2(\Lambda_X^2)+20\kt  L_X^4(\Lambda_X^2)-32L_X^6(\Lambda_X^2)}{|\bm{k}_T|^7 [\kt +4L_X^2(\Lambda_X^2)]^{5/2}}T_h(|\bm{k}_T|)\bigg]\,,
\end{align}
\begin{align}
\mathcal{D}_8 \left(x,\kt \right)=&\frac{1}{2P^+}\int \frac{d^2 \bm{l}_T}{(2\pi)^2} \frac{(\bm{l}_T \cdot \bm{k}_T)^3}{\bm{k}_T^6} \frac{1}{[\bm{l}_T^2+L_X^2(\Lambda_X^2)]^2} \frac{1}{[(\bm{l}_T+\bm{k}_T)^2+L_X^2(\Lambda_X^2)]^2}\notag\\
=&-\frac{1}{8\pi P^+} \bigg[ \frac{2+11L_X^2(\Lambda_X^2)/\kt +24L_X^4(\Lambda_X^2)/\ktquad }{2L_X^2(\Lambda_X^2)[\kt +4L_X^2(\Lambda_X^2)]^2}- \frac{\ktquad +22\kt  L_X^2(\Lambda_X^2)+48L_X^4(\Lambda_X^2)}{|\bm{k}_T|^5 [\kt +4L_X^2(\Lambda_X^2)]^{5/2}}T_h(|\bm{k}_T|)\bigg]\,,
\end{align}
\begin{align}
\mathcal{D}_9 \left(x,\kt \right)=&\frac{1}{2P^+}\int \frac{d^2 \bm{l}_T}{(2\pi)^2} \frac{(\bm{l}_T \cdot \bm{k}_T)^4}{\bm{k}_T^8} \frac{1}{[\bm{l}_T^2+L_X^2(\Lambda_X^2)]^2} \frac{1}{[(\bm{l}_T+\bm{k}_T)^2+L_X^2(\Lambda_X^2)]^2}\notag\\
=&\frac{1}{8\pi P^+} \bigg[ \frac{4+27L_X^2(\Lambda_X^2)/\kt +56L_X^4(\Lambda_X^2)/\ktquad }{4L_X^2(\Lambda_X^2)[\kt +4L_X^2(\Lambda_X^2)]^2}- \frac{2\ktquad +23\kt  L_X^2(\Lambda_X^2)+48L_X^4(\Lambda_X^2)}{|\bm{k}_T|^5 [\kt +4L_X^2(\Lambda_X^2)]^{5/2}}T_h(|\bm{k}_T|)\bigg]\,,
\end{align}
\begin{align}
\mathcal{D}_{10} \left(x,\kt \right)=&\frac{1}{2P^+}\int \frac{d^2 \bm{l}_T}{(2\pi)^2} \frac{(\bm{l}_T \cdot \bm{k}_T)^3}{\bm{k}_T^6} \frac{\bm{l}_T^2}{\kt } \frac{1}{[\bm{l}_T^2+L_X^2(\Lambda_X^2)]^2} \frac{1}{[(\bm{l}_T+\bm{k}_T)^2+L_X^2(\Lambda_X^2)]^2}\notag\\
=&-\frac{1}{8\pi P^+} \bigg[ \frac{2+13L_X^2(\Lambda_X^2)/\kt +17L_X^4(\Lambda_X^2)/\ktquad -24L_X^6(\Lambda_X^2)/\bm{k}_T^6}{2L_X^2(\Lambda_X^2)[\kt +4L_X^2(\Lambda_X^2)]^2}\notag\\
&- \frac{\bm{k}_T^6+9\ktquad L_X^2(\Lambda_X^2)+2\kt  L_X^4(\Lambda_X^2)-48L_X^6(\Lambda_X^2)}{|\bm{k}_T|^7 [\kt +4L_X^2(\Lambda_X^2)]^{5/2}}T_h(|\bm{k}_T|)\bigg]\,,
\end{align}
\begin{align}
\mathcal{D}_{11} \left(x,\kt \right)=&\frac{1}{2P^+}\int \frac{d^2 \bm{l}_T}{(2\pi)^2} \frac{(\bm{l}_T \cdot \bm{k}_T)}{\bm{k}_T^2} \frac{\bm{l}_T^4}{\ktquad } \frac{1}{[\bm{l}_T^2+L_X^2(\Lambda_X^2)]^2} \frac{1}{[(\bm{l}_T+\bm{k}_T)^2+L_X^2(\Lambda_X^2)]^2}\notag\\
=&-\frac{1}{8\pi P^+} \bigg[ \frac{1+6L_X^2(\Lambda_X^2)/\kt +5L_X^4(\Lambda_X^2)/\ktquad -18L_X^6(\Lambda_X^2)/\bm{k}_T^6}{L_X^2(\Lambda_X^2)[\kt +4L_X^2(\Lambda_X^2)]^2}\notag\\
&+ \frac{24L_X^4(\Lambda_X^2)(\kt +3L_X^2(\Lambda_X^2))}{|\bm{k}_T|^7 [\kt +4L_X^2(\Lambda_X^2)]^{5/2}}T_h(|\bm{k}_T|)\bigg]\,,
\end{align}

\section{Full expressions of gluon TMDs}\label{appendix2}

In the following, we list the final expressions of the $\mathcal{C}_{ijk}^{[F],l}$ coefficients in Eqs.~\eqref{eq:F_IM} and~\eqref{eq:N_dy} for each twist-3 gluon TMDs $F^{\Im}$ and $\tilde{N}$, and for $l=1,...,8$, $i,j,k=1,2$. We note that $i,j$, and $k$ correspond to $g_i (k^2)$, $g_j (l^2)$, and $g_k ((k+l)^2)$, respectively.

\subsection{\texorpdfstring{$G_3^\perp$}{G3perp}}
\begin{table}[H]
    \centering
    \caption{Coefficients functions of $G_3^\perp$ for $ i j k = \{111,112\} $.}
    \renewcommand{\arraystretch}{2}
    \begin{tabular}{|c|c|c|}
        \hline
        $C_{ijk}^{[G_3^\perp], l}$ & $i j k = 111$ & $i j k = 112$ \\
        \hline
        $l = 1$ & 0 & 0 \\
        \hline
        $l = 2$ &\begin{tabular}{c} $-\frac{24}{x}\big[\kt  (x-2)+M^2(x^3-2x^2+3x-2)$\\$+2xM M_X-M_X^2 (x-2)\big]$\end{tabular} &\begin{tabular}{c} $\frac{12}{M(x-1)}\big[\kt  M_X x+2M^3(x-1)^2+M^2M_X(x^3-2x^2+3x-2)$\\$+2M M_X^2(x-1)-M_X^3(x-2)]$\end{tabular} \\ 
        \hline
        $l = 3$ & 0 & $\frac{24M_X}{M(x-1)}\kt $ \\
        \hline
        $l = 4$ & $\frac{24(2-x)}{x} \kt $ & $\frac{12(M+M_X)}{M}\kt $ \\ 
        \hline
        $l = 5$ & 0 & $\frac{12(M(x-1)+M_X)}{M(x-1)} \kt $ \\
        \hline
        $l = 6$ & 0 & 0 \\
        \hline
        $l = 7$ & 0 & 0 \\
        \hline
        $l = 8$ & 0 & 0 \\
        \hline
    \end{tabular}
\end{table}

\begin{table}[H]
    \centering
    \caption{Coefficients functions of $G_3^\perp$ for $ i j k = \{121,122\} $.}
    \renewcommand{\arraystretch}{2}
    \begin{tabular}{|c|c|c|}
        \hline
        $C_{ijk}^{[G_3^\perp], l}$ & $i j k = 121$ & $i j k = 122$ \\
        \hline
        $l = 1$ & 0 & 0 \\
        \hline
        $l = 2$ &$-\frac{24(M+M_X)(\kt +(M(x-1)+M_X)^2)}{M}$ &\begin{tabular}{c} $\frac{6}{M^2(x-1)}[\ktquad +2\kt (M^2(x-1)x+2MM_X(x-1)+M_X^2 x) $\\$+(M(x-1)+M_X)^2(M^2(x^2-1)+2MM_X (x-1)+M_X^2 (2x-1))]$\end{tabular} \\
        \hline
        $l = 3$ & $-\frac{24M_X\kt }{M}$ & $\frac{12\kt [\kt  +M^2 (x-1)^2 +MM_X (x-1)x+M_X^2 (x^2+x-1)]}{M^2 (x-1)x}$ \\
        \hline
        $l = 4$ & $-24\kt $&$\frac{6[\kt +M^2(x^2-1)+4MM_X (x-1)+M_X^2]\kt }{M^2(x-1)}$ \\
        \hline
        $l = 5$ & $-\frac{12(M(x-1)+M_X)\kt }{M}$ & $\frac{6\kt  [\kt  (x-4)-2M^2 (x-1)^2 -M M_X x(x^2+x-2)-M_X^2 (x^2+2x-2)]}{M^2 (1-x)x}$ \\
        \hline
        $l = 6$ & 0 & $\frac{6(x-2)\ktquad }{M^2(1-x)x}$ \\
        \hline
        $l = 7$ & 0 & 0 \\
        \hline
        $l = 8$ & 0 & 0 \\
        \hline
    \end{tabular}
\end{table}

\begin{table}[H]
    \centering
    \caption{Coefficients functions of $G_3^\perp$ for $ i j k = \{211,212\} $.}
    \renewcommand{\arraystretch}{2}
    \begin{tabular}{|c|c|c|}
        \hline
        $C_{ijk}^{[G_3^\perp], l}$ & $i j k = 211$ & $i j k = 212$ \\
        \hline
        $l = 1$ & 0 & 0 \\
        \hline
        $l = 2$ &\begin{tabular}{c}$\frac{12}{M(x-1)}[\kt  M_X x+2M^3 (x-1)^2+M^2 M_X $\\$\times(x^3-2x^2+3x-2)+2MM_X^2 (x-1)-M_X^3 (x-2)]$ \end{tabular}&\begin{tabular}{c}$\frac{6}{M^2 (1-x)x}\{2\ktquad +\kt  [M^2 (x-1)(3x-2)+2MM_X x$\\$+M_X^2(x^2+3x-2)]+x(M(x-1)+M_X)[M x^2 (M^2+M_X^2)$\\$+M_X x(M^2-M_X^2)-(M-M_X)(M+M_X)^2]\}$ \end{tabular} \\
        \hline
        $l = 3$ & 0 & $\frac{12\kt  [2\kt + M^2 (x-1)^2+MM_X x+M_X^2 (x-1)]}{M^2 (1-x)x}$ \\
        \hline
        $l = 4$ & $\frac{12(M_X-M)\kt }{M}$&$\frac{6\kt  [2\kt + M^2 (x-1)x+M_X^2 x (x+1)]}{M^2 (1-x)x}$ \\
        \hline
        $l = 5$ & 0 &$\frac{6\kt  [2\kt + x(M+M_X)(M(x-1)+M_X)]}{M^2 (1-x)x}$ \\
        \hline
        $l = 6$ & 0 & 0 \\
        \hline
        $l = 7$ & 0 & 0 \\
        \hline
        $l = 8$ & 0 & 0 \\
        \hline
    \end{tabular}
\end{table}

\begin{table}[H]
    \centering
    \caption{Coefficients functions of $G_3^\perp$ for $ i j k = \{221,222\} $.}
    \renewcommand{\arraystretch}{2}
    \begin{tabular}{|c|c|c|}
        \hline
        $C_{ijk}^{[G_3^\perp], l}$ & $i j k = 221$ & $i j k = 222$ \\
        \hline
        $l = 1$ & 0 & 0 \\
        \hline
        $l = 2$ &\begin{tabular}{c} $\frac{6}{M^2 (x-1)} (\kt +(M(x-1)+M_X)^2)$\\$\times (\kt +M^2 x^2+2M_X x(M+M_X)-(M+M_X)^2)$ \end{tabular}&\begin{tabular}{c}$\frac{6}{M^3(1-x)}(M+M_X)(\kt +(M(x-1)+M_X)^2)$\\$\times(\kt +x(M^2(x-1)+M_X^2))$\end{tabular} \\
        \hline
        $l = 3$ & $\frac{12\kt (\kt +M^2(x-1)^2+MM_X x+M_X^2 (x-1))}{xM^2}$ & $\frac{6\kt }{M^3 (1-x)}(M+M_X)(\kt +M^2(x-1)^2+MM_X(x-1)+M_X^2 x)$ \\
        \hline
        $l = 4$ & $\frac{6\kt (\kt  +M^2(x^2-1)+M_X^2)}{M^2(x-1)}$&$\frac{6\kt }{M^3 (1-x)}(M+M_X)(\kt +M^2 (x-1)x+MM_X (x-1)+M_X^2)$ \\
        \hline
        $l = 5$ & $\frac{6\kt  (2\kt +x(M+M_X)(M(x-1)+M_X))}{xM^2}$ &\begin{tabular}{c} $\frac{3}{M^3(1-x)}[\kt (3M+M_X)+(M(x-1)+M_X)$\\$\times(2M^2 (x-1)+MM_X x+M_X^2(x+2)) ]$ \end{tabular}\\
        \hline
        $l = 6$ & 0 & $\frac{3\ktquad  (M_X-M)}{M^3(x-1)}$ \\
        \hline
        $l = 7$ & 0 & 0 \\
        \hline
        $l = 8$ & 0 & 0 \\
        \hline
    \end{tabular}
\end{table}

\subsection{\texorpdfstring{$\Delta G_{3L}^\perp$}{DeltaG3Lperp}}
\begin{table}[H]
    \centering
    \caption{Coefficients functions of $\Delta G_{3L}^\perp$ for $ i j k = \{111,112\} $.}
    \renewcommand{\arraystretch}{2}
    \begin{tabular}{|c|c|c|}
        \hline
        $C_{ijk}^{[\Delta G_{3L}^\perp], l}$ & $i j k = 111$ & $i j k = 112$ \\
        \hline
        $l = 1$ & 0 & $24(\kt -M^2(x-1)^2+M_X^2)$ \\
        \hline
        $l = 2$ &$24(\kt -M^2(x^2-1)-2MM_X x-M_X^2)$&\begin{tabular}{c} $\frac{12}{M(x-1)}[\kt (4M(x-1)-M_X(x-2))$\\$+(M(x-1)+M_X)(2M^2(x-1)+MM_X(x^2+x-2)+M_X^2 x)]$ \end{tabular} \\
        \hline
        $l = 3$ & 0 & $\frac{24\kt  (2M(x-1)+M_X)}{M(x-1)}$ \\
        \hline
        $l = 4$ & $24\kt $&$-\frac{12(M+M_X)}{M}\kt $ \\
        \hline
        $l = 5$ & 0 &$\frac{12\kt  (M(x-1)+M_X)}{M(x-1)}$ \\
        \hline
        $l = 6$ & 0 & 0 \\
        \hline
        $l = 7$ & 0 & 0 \\
        \hline
        $l = 8$ & 0 & 0 \\
        \hline
    \end{tabular}
\end{table}

\begin{table}[H]
    \centering
    \caption{Coefficients functions of $\Delta G_{3L}^\perp$ for $ i j k = \{121,122\} $.}
    \renewcommand{\arraystretch}{2}
    \begin{tabular}{|c|c|c|}
        \hline
        $C_{ijk}^{[\Delta G_{3L}^\perp], l}$ & $i j k = 121$ & $i j k = 122$ \\
        \hline
        $l = 1$ & 0 & 0 \\
        \hline
        $l = 2$ &$-\frac{24}{M}[\kt (M(2x-1)+M_X)+(M+M_X)(M(x-1)+M_X)^2]$&\begin{tabular}{c} $\frac{6}{M^2(x-1)}[\ktquad +2\kt  x(M^2(x-1)$\\$+2M M_X(x-1)+M_X^2)+(M(x-1)+M_X)^2$\\$\times(M^2 x^2+2M_X x(M+M_X)-(M+M_X)^2)]$ \end{tabular} \\
        \hline
        $l = 3$ & $-\frac{24\kt (2M(x-1)+M_X)}{M}$ &\begin{tabular}{c} $\frac{12\kt }{M^2(x-1)x}[\kt  (2x-1)+M^2 (x-1)^2$\\$+M M_X(x-1)x(2x+1)+M_X^2(x^2+x-1)]$\end{tabular} \\
        \hline
        $l = 4$ & $-24\kt $&\begin{tabular}{c}$\frac{6\kt }{M^2 (1-x)x}[\kt  (x-2)-M^2 (x-1)(x^2-x+2)$\\$+2MM_X(x-1)x+M_X^2(x-2)]$\end{tabular} \\
        \hline
        $l = 5$ & $-\frac{12\kt (M(x-1)+M_X)}{M}$ &$\frac{6\kt  (3\kt +M_X x(M(x-1)+M_X))}{M^2(x-1)}$ \\
        \hline
        $l = 6$ & 0 & $\frac{6\ktquad }{M^2(x-1)}$ \\
        \hline
        $l = 7$ & 0 & 0 \\
        \hline
        $l = 8$ & 0 & 0 \\
        \hline
    \end{tabular}
\end{table}

\begin{table}[H]
    \centering
    \caption{Coefficients functions of $\Delta G_{3L}^\perp$ for $ i j k = \{211,212\} $.}
    \renewcommand{\arraystretch}{2}
    \begin{tabular}{|c|c|c|}
        \hline
        $C_{ijk}^{[\Delta G_{3L}^\perp], l}$ & $i j k = 211$ & $i j k = 212$ \\
        \hline
        $l = 1$ & $-24(\kt -M^2(x-1)^2+M_X^2)$ & 0 \\
        \hline
        $l = 2$ &\begin{tabular}{c} $-\frac{12}{M(1-x)} [\kt (2M(x-1)+M_X(x-2))$\\$-x(M(x-1)+M_X)(2M^2(x-1)$\\$+MM_X (x-1)+M_X^2)]$ \end{tabular} & \begin{tabular}{c} $\frac{6}{M^2(1-x)}(M+M_X)[\kt  (3M(x-1)-M_X(x-3))$\\$+(M(x-1)+M_X)(M^2(x^2-1)$\\$+MM_X(x-1)x+M_X^2(x+1))]$  \end{tabular} \\
        \hline
        $l = 3$ & 0 &$\frac{12}{M^2(1-x)x} [\kt +M^2(x^2-1)+MM_X x(2x-1)+M_X^2(x+1)]$ \\
        \hline
        $l = 4$ & $-\frac{12\kt (M+M_X)}{M}$ &\begin{tabular}{c}$\frac{6\kt }{M^2(x-1)x}[2\kt -(x-1)(M^2 (x-2)$\\$-2M M_X x-M_X^2 x)+2M_X^2]$\end{tabular} \\
        \hline
        $l = 5$ & 0 & $\frac{6\kt  (M+M_X)(M(x-1)+M_X)}{M^2(1-x)}$ \\
        \hline
        $l = 6$ & 0 & 0 \\
        \hline
        $l = 7$ & 0 & 0 \\
        \hline
        $l = 8$ & 0 & 0 \\
        \hline
    \end{tabular}
\end{table}

\begin{table}[H]
    \centering
    \caption{Coefficients functions of $\Delta G_{3L}^\perp$ for $ i j k = \{221,222\} $.}
    \renewcommand{\arraystretch}{2}
    \begin{tabular}{|c|c|c|}
        \hline
        $C_{ijk}^{[\Delta G_{3L}^\perp], l}$ & $i j k = 221$ & $i j k = 222$ \\
        \hline
        $l = 1$ & 0 & 0 \\
        \hline
        $l = 2$ & \begin{tabular}{c}$\frac{6}{M^2(x-1)} [\ktquad +2\kt  x(M^2(x-1)$\\$+2MM_X(x-1)+M_X^2)$\\$+(M(x-1)+M_X)^2(M^2 x^2$\\$+2M_X x(M+M_X)-(M+M_X)^2)]$ \end{tabular} & \begin{tabular}{c}$\frac{6}{M^3(1-x)} (M+M_X) [\ktquad +\kt  (M^2(x-1)$\\$+2MM_X(x-1)x+M_X^2(x+1))$\\$+x(M(x-1)+M_X)^2 (M^2 (x-1)+M_X^2)]$ \end{tabular} \\
        \hline
        $l = 3$ &\begin{tabular}{c} $\frac{12\kt }{xM^2} [\kt +M^2(x^2-1)$\\$+MM_X x(2x-1)+M_X^2(x+1)]$\end{tabular} &\begin{tabular}{c} $\frac{6\kt }{M^3(1-x)} [\kt (M+3M_X)+M^3(x-1)^2$\\$+M^2M_X(x^2+x-2)+MM_X^2(2x^2-1)+M_X^3(x+2)]$\end{tabular} \\
        \hline
        $l = 4$ & \begin{tabular}{c}$\frac{6\kt }{M^2(1-x)x}[\kt (x-2)-M^2(x-1)(x^2-x+2)$\\$-2MM_X(x-1)x+M_X^2(x-2)] $ \end{tabular} &$\frac{6\kt }{M^3(1-x)} (M-M_X)(\kt -M^2(x-2)(x-1)+MM_X(x-1)+M_X^2)$ \\
        \hline
        $l = 5$ & $\frac{6\kt (M+M_X)(M(x-1)+M_X)}{M^2}$ & $\frac{3\kt }{M^3(1-x)}(M+M_X)(3\kt +M_X x(M(x-1)+M_X))$ \\
        \hline
        $l = 6$ & 0 & $\frac{3\ktquad  (M+M_X)}{M^3 (1-x)}$ \\
        \hline
        $l = 7$ & 0 & 0 \\
        \hline
        $l = 8$ & 0 & 0 \\
        \hline
    \end{tabular}
\end{table}

\subsection{\texorpdfstring{$\Delta G_{3T}$}{DeltaG3T}}

\begin{table}[H]
    \centering
    \caption{Coefficients functions of $\Delta G_{3T}$ for $ i j k = \{111,112\} $.}
    \renewcommand{\arraystretch}{2}
    \begin{tabular}{|c|c|c|}
        \hline
        $C_{ijk}^{[\Delta G_{3T}], l}$ & $i j k = 111$ & $i j k = 112$ \\
        \hline
        $l = 1$ & 0 & \begin{tabular}{c} $\frac{12}{M^2(x-1)x}[\ktquad (2-x)+2\kt (M_X^2$\\$-M^2(x-1)^2(x+1))+x(M_X^2+M^2(x-1)^2)^2]$ \end{tabular} \\
        \hline
        $l = 2$ &$-\frac{48\kt (M(x-1)+M_X)}{M}$& \begin{tabular}{c} $\frac{12\kt }{M^2 (1-x)x} [\kt (2x-5)+(x-1)(M^2(x-1)(x+3)$\\$+MM_X(1-2x)x-2M_X^2 x)-3M_X^2]$  \end{tabular} \\
        \hline
        $l = 3$ & 0 & \begin{tabular}{c} $\frac{6\kt }{M^2 (x-1)x} [\kt (x^2-3x+6)+(x-1)(M^2(x-1)(x^2-x+2)$\\$+MM_X x(x+2)-M_X^2(x-2))]$   \end{tabular}\\
        \hline
        $l = 4$ & $-\frac{24\kt (M(x-1)+M_X)}{M}$ &\begin{tabular}{c} $\frac{6\kt }{M^2 (x-1)x}[\kt (-x^2+x+2)+x(M^2(x-1)^3$\\$+3MM_X x (x-1)+M_X^2(x+1))]$ \end{tabular}\\
        \hline
        $l = 5$ & 0 & $\frac{12\ktquad }{M^2 (x-1)x}$ \\
        \hline
        $l = 6$ & 0 & 0 \\
        \hline
        $l = 7$ & 0 & 0 \\
        \hline
        $l = 8$ & 0 & 0 \\
        \hline
    \end{tabular}
\end{table}

\begin{table}[H]
    \centering
    \caption{Coefficients functions of $\Delta G_{3T}$ for $ i j k = \{121,122\} $.}
    \renewcommand{\arraystretch}{2}
    \begin{tabular}{|c|c|c|}
        \hline
        $C_{ijk}^{[\Delta G_{3T}], l}$ & $i j k = 121$ & $i j k = 122$ \\
        \hline
        $l = 1$ & 0 & 0 \\
        \hline
        $l = 2$ &$-\frac{24\kt  }{xM^2}[\kt +(M(x-1)+M_X)(M(1-x^2)+M_X)]$ &-$\frac{6\kt }{M^2}[\kt +(M(x-1)+M_X)(M(x-1)+M_X(2x-1))]$ \\
        \hline
        $l = 3$ &\begin{tabular}{c} $\frac{6\kt }{xM^2}[2\kt (x-3)+(M(x-1)+M_X)$\\$\times(M(x^2+2x-2)+2M_X(x-1))]$ \end{tabular}& \begin{tabular}{c} $\frac{3\kt }{4M^3(1-x)}[\kt (M(28x-26)+M_X x)$\\$x(M^3(x-1)^2+M^2M_X(x-1)(11x-13)$\\$+MM_X^2(11x-9)+M_X^3)]$  \end{tabular}\\
        \hline
        $l = 4$ & $-\frac{6\kt }{M^2}[2\kt -(M(x-1)+M_X)(xM-2(M+M_X))]$ &\begin{tabular}{c} $\frac{3\kt }{4M^3(x-1)}[\kt (M(6-4x)+M_X(x-8))$\\$+M^3(x-1)^2x+M^2M_X(3x^3-11x+8)$\\$+MM_X^2x (3x-1)+M_X^3(x-8)]$   \end{tabular}\\
        \hline
        $l = 5$ & $-\frac{12\ktquad }{xM^2}$ & $\frac{3\ktquad (M(x-1)(9x+8)+6M_X x+8M_X)}{4M^3(1-x)}$ \\
        \hline
        $l = 6$ & 0 & $\frac{3\ktquad (3M(x-1)x+M_X(3x+4))}{4M^3(1-x)}$ \\
        \hline
        $l = 7$ & 0 & $\frac{3\ktquad  (M(x-1)(3x-8)+M_X(3x-4))}{4M^3(x-1)}$ \\
        \hline
        $l = 8$ & 0 & $\frac{3\ktquad (3M(x-1)(3x-8)+6M_X x-8M_X)}{4M^3(x-1)}$ \\
        \hline
    \end{tabular}
\end{table}

\begin{table}[H]
    \centering
    \caption{Coefficients functions of $\Delta G_{3T}$ for $ i j k = \{211,212\} $.}
    \renewcommand{\arraystretch}{2}
    \begin{tabular}{|c|c|c|}
        \hline
        $C_{ijk}^{[\Delta G_{3T}], l}$ & $i j k = 211$ & $i j k = 212$ \\
        \hline
        $l = 1$ & \begin{tabular}{c} $\frac{12}{M^2(1-x)x}[\ktquad (2-x)+2\kt $\\$\times(M_X^2-M^2(x-1)^2(x+1))$\\$+x(M_X^2-M^2(x-1)^2)^2]$ \end{tabular}& 0 \\
        \hline
        $l = 2$ &\begin{tabular}{c} $\frac{12\kt }{M^2(x-1)x}[\kt (2x-3)+(x-1)$\\$\times(M^2(x-1)(3x+1)+MM_X x(2x-1)$\\$+2M_X^2 x)-M_X^2]$ \end{tabular}&\begin{tabular}{c} $\frac{12\kt }{M^3(1-x)}[\kt M+(M(x-1)+M_X)M^2(1-x)$\\$+MM_X x+M_X^2(x-1)]$ \end{tabular}\\
        \hline
        $l = 3$ & 0 & \begin{tabular}{c}$\frac{3\kt }{4M^3(1-x)}[\kt (M(x^2-9x+24)+2M_X(x-4))$\\$-2(M(x-1)+M_X)(M^2(x-1)(x+4)$\\$+2MM_X(x-1)x-M_X^2(x-4))]$   \end{tabular}\\
        \hline
        $l = 4$ & $\frac{12\kt (\kt +x(M+M_X)(M(x-1)+M_X))}{xM^2}$ &\begin{tabular}{c}$\frac{3\kt }{4M^3(1-x)}[\kt (M(9-x)x-2M_X(x-4))$\\$+2x(M(x-1)+M_X)(M^2(x-1)+2MM_X(x+1)+3M_X^2)]$ \end{tabular}\\
        \hline
        $l = 5$ & 0 & $\frac{6\ktquad }{M^2(1-x)}$ \\
        \hline
        $l = 6$ & 0 & 0 \\
        \hline
        $l = 7$ & 0 & 0 \\
        \hline
        $l = 8$ & 0 & 0 \\
        \hline
    \end{tabular}
\end{table}

\begin{table}[H]
    \centering
    \caption{Coefficients functions of $\Delta G_{3T}$ for $ i j k = \{221,222\} $.}
    \renewcommand{\arraystretch}{2}
    \begin{tabular}{|c|c|c|}
        \hline
        $C_{ijk}^{[\Delta G_{3T}], l}$ & $i j k = 221$ & $i j k = 222$ \\
        \hline
        $l = 1$ & 0 & 0 \\
        \hline
        $l = 2$ &\begin{tabular}{c}$-\frac{6\kt }{M^2}[\kt +(M(x-1)+M_X)$\\$\times(M(x-1)+M_X(2x-1))]$\end{tabular} &\begin{tabular}{c} $\frac{6\kt }{M^4(x-1)x}[\ktquad +\kt (M^2(x^2-1)+MM_X(x-1)x$\\$+M_X^2(x+1))+x(M^4(1-x)^3+M^3M_X(x-1)^2x$\\$+M^2M_X^2(x^3-2x^2+3x-2)+MM_X^3(x-1)x+M_X^4)]$  \end{tabular} \\
        \hline
        $l = 3$ & \begin{tabular}{c} $\frac{3\kt }{4M^3(x-1)}[\kt (2M(3x^2+x-4)+M_X(8-9x))$\\$-M^3(x-1)^2(13x-8)-M^2M_X(x-1)(3x^2+9x-8)$\\$-MM_X^2(x-1)(13x-8)+M_X^3(8-9x)]$ \end{tabular}& \begin{tabular}{c} $-\frac{3\kt }{16M^4(x-1)^2 x} [\ktquad (7x^3-12x^2-50x+56)$\\$+\kt (M^2(3x^5-4x^4-40x^3+57x^2+8x-24)$\\$+MM_X x(-41x^2+81x-40)+2M_X^2(-22x^2+11x+12))$\\$-x(M^4(x-1)^3(8x^2-13x+8)+M^3M_X(x-1)^2$\\$\times x(24x-23)+M^2M_X^2(17x^4-40x^3+44x^2-37x+16)$\\$+MM_X^3 x(24x^2-47x+23)+M_X^4(7x^2-8))]$    \end{tabular}\\
        \hline
        $l = 4$ &\begin{tabular}{c} $\frac{3\kt }{4M^3(x-1)}[\kt (9M_X-6M(x-1))$\\$+5M^3(x-1)^2-M^2M_X(x-1)(5x-9)$\\$+5MM_X^2(x-1)+9M_X^3]$\end{tabular} &\begin{tabular}{c}$-\frac{3\kt }{16M^4(x-1)^2 x}[\kt (-7x^3+12x^2-30x+24)$\\$-\kt (M^2(3x^5-4x^4+8x^3+9x^2-40x+24)$\\$+MM_X x(7x^2-15x+8)+2M_X^2(2x^2+11x-12))$\\$x(M^4(x-1)^3(8x^2+3x-8)+M^3M_X(x-1)^2x(8x-7)$\\$+M^2M_X^2(x^4+8x^3-36x^2+43x-16)$\\$+MM_X^3x(8x^2-15x+7)+M_X^4(7x^2-16x+8))]$ \end{tabular}\\
        \hline
        $l = 5$ & $\frac{6\ktquad }{M^2}$ & \begin{tabular}{c} $\frac{3\ktquad }{16M^4(x-1)x}[\kt (-3x^2+16x+48)+2M^2(3x^4-4x^3$\\$+16x^2-4x-8)+4MM_X x(4x^2+3x-2)$\\$+M_X^2(13x^2+22x+16)]$   \end{tabular}\\
        \hline
        $l = 6$ & 0 &\begin{tabular}{c} $\frac{3\ktquad }{8M^4(x-1)x}[\kt (-2x^2+6x+4)+x(M^2(x^3-x^2+4x-4)$\\$+MM_Xx(3x+2)+2M_X^2(x+2))]$ \end{tabular}\\
        \hline
        $l = 7$ & 0 &\begin{tabular}{c} $\frac{3\ktquad }{8M^4(1-x)x}(x-2)[x(M^2(x^2+x-2)$\\$+3MM_X x+2M_X^2)-2\kt (x-1)]$\end{tabular} \\
        \hline
        $l = 8$ & 0 &\begin{tabular}{c} $\frac{3\ktquad }{16M^4(x-1)x}[\kt (3x^2-14x+16)$\\$+x(M^2(-6x^3+8x^2+22x-24)$\\$-4MM_X(4x^2-9x+2)-13M_X^2(x-2))]$ \end{tabular}\\
        \hline
    \end{tabular}
\end{table}

\subsection{\texorpdfstring{$\Delta G_{3T}^\perp$}{DeltaG3Tperp}}

\begin{table}[H]
    \centering
    \caption{Coefficients functions of $\Delta G_{3T}^\perp$ for $ i j k = \{111,112\} $.}
    \renewcommand{\arraystretch}{2}
    \begin{tabular}{|c|c|c|}
        \hline
        $C_{ijk}^{[\Delta G_{3T}^\perp], l}$ & $i j k = 111$ & $i j k = 112$ \\
        \hline
        $l = 1$ & 0& $\frac{48(\kt +M^2(x-1)+M_X^2)}{x}$ \\
        \hline
        $l = 2$ &0&$\frac{24(5\kt +3M^2(x-1)+MM_X x+3M_X^2)}{x}$\\
        \hline
        $l = 3$ & 0 &$\frac{24(3\kt +(M+M_X)(M(x-1)+M_X))}{x}$\\
        \hline
        $l = 4$ & 0 &$\frac{24\kt }{x}$\\
        \hline
        $l = 5$ & 0 & $\frac{24\kt }{x}$ \\
        \hline
        $l = 6$ & 0 & 0 \\
        \hline
        $l = 7$ & 0 & 0 \\
        \hline
        $l = 8$ & 0 & 0 \\
        \hline
    \end{tabular}
\end{table}

\begin{table}[H]
    \centering
    \caption{Coefficients functions of $\Delta G_{3T}^\perp$ for $ i j k = \{121,122\} $.}
    \renewcommand{\arraystretch}{2}
    \begin{tabular}{|c|c|c|}
        \hline
        $C_{ijk}^{[\Delta G_{3T}^\perp], l}$ & $i j k = 121$ & $i j k = 122$ \\
        \hline
        $l = 1$ & 0& 0 \\
        \hline
        $l = 2$ &$\frac{48(1-x)(\kt +(M+M_X)(M(x-1)+M_X))}{x}$&\begin{tabular}{c} $\frac{12}{M}[\kt (M+2M_X)+(M(x-1)+M_X)$\\$\times(M^2(x-1)+MM_X+2M_X^2)]$   \end{tabular}\\
        \hline
        $l = 3$ & $\frac{24(1-x)(3\kt +(M+M_X)(M(x-1)+M_X))}{x}$&\begin{tabular}{c} $\frac{3}{M}[\kt (M(x+10)+24M_X)$\\$8M_X(M+M_X)(M(x-1)+M_X)]$    \end{tabular}\\
        \hline
        $l = 4$ & 0 &$\frac{3(6-x)\kt }{2}$\\
        \hline
        $l = 5$ &  $\frac{24\kt  (1-x)}{x}$&$\frac{3\kt (5Mx+8M_X)}{2M}$ \\
        \hline
        $l = 6$ & 0 & 0 \\
        \hline
        $l = 7$ & 0 & 0 \\
        \hline
        $l = 8$ & 0 & $\frac{3\kt (8-5x)}{2}$ \\
        \hline
    \end{tabular}
\end{table}

\begin{table}[H]
    \centering
    \caption{Coefficients functions of $\Delta G_{3T}^\perp$ for $ i j k = \{211,212\} $.}
    \renewcommand{\arraystretch}{2}
    \begin{tabular}{|c|c|c|}
        \hline
        $C_{ijk}^{[\Delta G_{3T}^\perp], l}$ & $i j k = 211$ & $i j k = 212$ \\
        \hline
        $l = 1$ & $-\frac{48(\kt +M^2(x-1)+M_X^2)}{x}$& 0 \\
        \hline
        $l = 2$ & $-\frac{24(3\kt +(M-M_X)(M(x-1)-M_X))}{x}$ &$\frac{24}{M(1-x)}[\kt M_X +(M(x-1)+M_X)(M^2(x-1)+M_X^2)]$\\
        \hline
        $l = 3$ & 0 & \begin{tabular}{c}$\frac{3}{2M(x-1)}[\kt (M(x-1)(3x-8)-24M_X)$\\$-8(M(x-1)+M_X)(M^2(x-1)+M_X^2)]$ \end{tabular}\\
        \hline
        $l = 4$ &$-\frac{24\kt }{x}$&$\frac{3\kt (8-3x)}{2}$\\
        \hline
        $l = 5$ & 0 & $\frac{12M_X \kt }{M(1-x)}$ \\
        \hline
        $l = 6$ & 0 &0\\
        \hline
        $l = 7$ & 0 & 0 \\
        \hline
        $l = 8$ & 0 & 0 \\
        \hline
    \end{tabular}
\end{table}

\begin{table}[H]
    \centering
    \caption{Coefficients functions of $\Delta G_{3T}^\perp$ for $ i j k = \{221,222\} $.}
    \renewcommand{\arraystretch}{2}
    \begin{tabular}{|c|c|c|}
        \hline
        $C_{ijk}^{[\Delta G_{3T}^\perp], l}$ & $i j k = 221$ & $i j k = 222$ \\
        \hline
        $l = 1$ & 0& 0 \\
        \hline
        $l = 2$ & \begin{tabular}{c} $\frac{12}{M}[\kt (M+2M_X)$\\$+(M(x-1)+M_X)(M^2(x-1)+MM_X+2M_X^2)]$  \end{tabular} &\begin{tabular}{c} $\frac{12}{M^2(1-x)x}[\ktquad +\kt (M^2(x-1)+MM_X(x-1)x$\\$+M_X^2(x^2+1))+M_X x^2(M(x-1)+M_X)(M^2(x-1)+M_X^2)]$  \end{tabular}\\
        \hline
        $l = 3$ & \begin{tabular}{c} $-\frac{3}{2M}[\kt (9xM-8(M+3M_X))$\\$-8(M(x-1)+M_X)(M^2(x-1)+M_X^2)]$  \end{tabular} & \begin{tabular}{c} $\frac{3}{8M^2(x-1)x}[\ktquad (5x^2-2x-56)$\\$+\kt (M^2(x-1)(5x^3-5x^2+4x-24)$\\$+10MM_X(x-4)(x-1)x-M_X^2(43x^2+2x+24))$\\$-16M_X x^2(M(x-1)+M_X)(M^2(x-1)+M_X^2)]$  \end{tabular} \\
        \hline
        $l = 4$ &$\frac{27}{2}x\kt $&\begin{tabular}{c} $\frac{3\kt }{8M^2 (1-x)x}[\kt (5x^2-2x+24)$\\$+M^2(x-1)(5x^3-5x^2+4x+24)$\\$+2MM_X(x-1)x(5x+4)+M_X^2(5x^2-2x+24)]$  \end{tabular}\\
        \hline
        $l = 5$ & $\frac{12\kt M_X}{M}$ &\begin{tabular}{c}$\frac{3\kt }{8M^2(1-x)x}[\kt (x^2+6x+48)+2M^2(x^4-5x^2+12x-8)$\\$+2MM_X x(3x^2+x-4)+M_X^2(21x^2-10x+16)]$  \end{tabular}\\
        \hline
        $l = 6$ & 0 &$\frac{3\ktquad  (x^2-4x-4)}{4M^2 (x-1)x}$\\
        \hline
        $l = 7$ & 0 & $\frac{3\ktquad (x-2)^2}{4M^2(1-x)x}$ \\
        \hline
        $l = 8$ & 0 &\begin{tabular}{c} $\frac{3\kt }{8M^2(x-1)x}[\kt (x-2)(x+8)+x(2M^2(x^3-5x+4)$\\$+2MM_X(x-1)(3x-4)+5M_X^2(x-2))]$    \end{tabular}\\
        \hline
    \end{tabular}
\end{table}

\subsection{\texorpdfstring{$H_3^\perp$}{H3perp}}

\begin{table}[H]
    \centering
    \caption{Coefficients functions of $H_3^\perp$ for $ i j k = \{111,112\} $.}
    \renewcommand{\arraystretch}{2}
    \begin{tabular}{|c|c|c|}
        \hline
        $C_{ijk}^{[H_3^\perp], l}$ & $i j k = 111$ & $i j k = 112$ \\
        \hline
        $l = 1$ & 0& 0 \\
        \hline
        $l = 2$ & $\frac{24(\kt (x-2)-x(M(x-1)+M_X)^2)}{x}$ &$\frac{12(M+M_X)(\kt +(M(x-1)+M_X)^2)}{M}$\\
        \hline
        $l = 3$ & $-\frac{48\kt }{x}$ & $\frac{24\kt  (M+M_X)}{M}$\\
        \hline
        $l = 4$ &$24\kt $ &$12(x-2)\kt $\\
        \hline
        $l = 5$ & 0 & $\frac{12\kt  (M(x-1)+M_X)}{M}$ \\
        \hline
        $l = 6$ & 0 &0\\
        \hline
        $l = 7$ & 0 & 0 \\
        \hline
        $l = 8$ & 0 & 0 \\
        \hline
    \end{tabular}
\end{table}

\begin{table}[H]
    \centering
    \caption{Coefficients functions of $H_3^\perp$ for $ i j k = \{121,122\} $.}
    \renewcommand{\arraystretch}{2}
    \begin{tabular}{|c|c|c|}
        \hline
        $C_{ijk}^{[H_3^\perp], l}$ & $i j k = 121$ & $i j k = 122$ \\
        \hline
        $l = 1$ & 0& 0 \\
        \hline
        $l = 2$ & 0 &0\\
        \hline
        $l = 3$ & $-\frac{24\kt (M(x-1)+M_X)}{xM}$ & $\frac{12\kt (\kt +(M+M_X)(M(x-1)+M_X))}{xM^2}$\\
        \hline
        $l = 4$ &$\frac{24\kt  (M(x-1)+M_X)}{M}$ &$-\frac{6\kt (\kt +M^2(x^2-1)+2MM_X x+M_X^2)}{M^2}$\\
        \hline
        $l = 5$ & $\frac{12\kt  (M(x-1)+M_X)}{M}$ & $-\frac{6\kt (2\kt  (x-1)+Mx^2(M(x-1)+M_X))}{xM^2}$ \\
        \hline
        $l = 6$ & 0 &0\\
        \hline
        $l = 7$ & 0 & $\frac{6\ktquad (2-x)}{xM^2}$ \\
        \hline
        $l = 8$ & 0 & $\frac{12\ktquad }{xM^2}$ \\
        \hline
    \end{tabular}
\end{table}

\begin{table}[H]
    \centering
    \caption{Coefficients functions of $H_3^\perp$ for $ i j k = \{211,212\} $.}
    \renewcommand{\arraystretch}{2}
    \begin{tabular}{|c|c|c|}
        \hline
        $C_{ijk}^{[H_3^\perp], l}$ & $i j k = 211$ & $i j k = 212$ \\
        \hline
        $l = 1$ & 0& 0 \\
        \hline
        $l = 2$ & \begin{tabular}{c} $\frac{12}{xM}[\kt (M(3x-2)-M_X(x-2))$\\$+x(M+M_X)(M(x-1)+M_X)^2]$  \end{tabular} &\begin{tabular}{c} $-\frac{6}{xM^2}[2\ktquad +\kt (M^2(2x^2+x-2)$\\$+2MM_X x+M_X^2(x+2))$\\$+x(M+M_X)^2(M(x-1)+M_X)^2]$  \end{tabular}\\
        \hline
        $l = 3$ & $24\kt $ & $-\frac{12\kt (2\kt +x(xM^2+MM_X+M_X^2))}{xM^2}$\\
        \hline
        $l = 4$ &$-\frac{12\kt (M+M_X)}{M}$ &$\frac{6\kt (2-x)(M+M_X)}{M}$\\
        \hline
        $l = 5$ & 0 & $-\frac{6\kt (M+M_X)(M(x-1)+M_X)}{M^2}$ \\
        \hline
        $l = 6$ & 0 &0\\
        \hline
        $l = 7$ & 0 & 0 \\
        \hline
        $l = 8$ & 0 & $-\frac{12\ktquad }{xM^2}$ \\
        \hline
    \end{tabular}
\end{table}

\begin{table}[H]
    \centering
    \caption{Coefficients functions of $H_3^\perp$ for $ i j k = \{221,222\} $.}
    \renewcommand{\arraystretch}{2}
    \begin{tabular}{|c|c|c|}
        \hline
        $C_{ijk}^{[H_3^\perp], l}$ & $i j k =221$ & $i j k =222$ \\
        \hline
        $l = 1$ & 0& 0 \\
        \hline
        $l = 2$ & 0 &0\\
        \hline
        $l = 3$ & $-\frac{12\kt (\kt -xM(M(x-1)+M_X))}{xM^2}$ & $-\frac{6\kt (M+M_X)(M(x-1)+M_X)}{M^2}$\\
        \hline
        $l = 4$ &$-\frac{12\kt (M+M_X) (M(x-1)+M_X)}{M^2}$&$\frac{3\kt (M+M_X)}{M^3}[\kt +M^2(x^2-1)+2MM_X x+M_X^2]$ \\
        \hline
        $l = 5$  & $-\frac{6\kt  (2\kt +x(M+M_X)(M(x-1)+M_X))}{xM^2}$&\begin{tabular}{c} $\frac{3\kt }{xM^3} [2\kt (M+M_X(1+x))$\\$+Mx^2(M+M_X)(M(x-1)+M_X)]$  \end{tabular}\\
        \hline
        $l = 6$ & 0 &0\\
        \hline
        $l = 7$ & 0 & $\frac{3\ktquad (M_X-M)}{M^3}$ \\
        \hline
        $l = 8$ & 0 & $-\frac{6\ktquad (M+M_X)}{xM^3}$ \\
        \hline
    \end{tabular}
\end{table}

\subsection{\texorpdfstring{$\Delta H_{3L}^\perp$}{DeltaH3Lperp}}

\begin{table}[H]
    \centering
    \caption{Coefficients functions of $\Delta H_{3L}^\perp$ for $ i j k = \{111,112\} $.}
    \renewcommand{\arraystretch}{2}
    \begin{tabular}{|c|c|c|}
        \hline
        $C_{ijk}^{[\Delta H_{3L}^\perp], l}$ & $i j k = 111$ & $i j k = 112$ \\
        \hline
        $l = 1$ & 0& $\frac{48\kt  (x-1)}{x}$ \\
        \hline
        $l = 2$ & $-\frac{12(\kt (5x-6)-x(M(x-1)+M_X)^2)}{x}$ &\begin{tabular}{c} $\frac{6}{xM}[\kt (5x(M+M_X)-2(5M+M_X)+4Mx^2)$\\$-x(M+M_X)(M(x-1)+M_X)^2]$   \end{tabular}\\
        \hline
        $l = 3$ & 0 & $\frac{18\kt  (M(x-1)+M_X)}{M}$\\
        \hline
        $l = 4$ &$\frac{12(2-x)\kt }{x}$ &$-\frac{6\kt (M(x^2-2x+2)+2M_X)}{xM}$\\
        \hline
        $l = 5$ & 0 & 0 \\
        \hline
        $l = 6$ & 0 &0\\
        \hline
        $l = 7$ & 0 & 0 \\
        \hline
        $l = 8$ & 0 & 0 \\
        \hline
    \end{tabular}
\end{table}

\begin{table}[H]
    \centering
    \caption{Coefficients functions of $\Delta H_{3L}^\perp$ for $ i j k = \{121,122\} $.}
    \renewcommand{\arraystretch}{2}
    \begin{tabular}{|c|c|c|}
        \hline
        $C_{ijk}^{[\Delta H_{3L}^\perp], l}$ & $i j k = 121$ & $i j k = 122$ \\
        \hline
        $l = 1$ & 0& 0 \\
        \hline
        $l = 2$ & $\frac{48\kt (1-x)(M(x-1)+M_X)}{xM}$ &$\frac{12\kt }{xM^2}[\kt +M^2(x-1)^2+2MM_X(x-1)x+M_X^2(2x-1)]$\\
        \hline
        $l = 3$ & $-\frac{12\kt (M(x-1)+M_X)}{M}$ &$\frac{3\kt }{xM^2}[\kt (3x-2)+x(M^2(-3x^2+8x-5)+2MM_Xx+5M_X^2)]$ \\
        \hline
        $l = 4$ &$\frac{12\kt (M(x-1)+M_X)}{xM}$ &$-\frac{3\kt }{xM^2}[\kt (x-6)+M^2(-x^3+2x^2+x-2)+2MM_Xx+M_X^2(x+2)]$\\
        \hline
        $l = 5$ & $-\frac{6\kt (M(x-1)+M_X)}{M}$ & $\frac{3\kt (2\kt +(M+M_X)(M(x-1)+M_X))}{M^2}$ \\
        \hline
        $l = 6$ & 0 &0\\
        \hline
        $l = 7$ & 0 & 0 \\
        \hline
        $l = 8$ & 0 & 0 \\
        \hline
    \end{tabular}
\end{table}

\begin{table}[H]
    \centering
    \caption{Coefficients functions of $\Delta H_{3L}^\perp$ for $ i j k = \{211,212\} $.}
    \renewcommand{\arraystretch}{2}
    \begin{tabular}{|c|c|c|}
        \hline
        $C_{ijk}^{[\Delta H_{3L}^\perp], l}$ & $i j k = 211$ & $i j k = 212$ \\
        \hline
        $l = 1$ & $\frac{48\kt  (1-x)}{x}$& 0 \\
        \hline
        $l = 2$ & \begin{tabular}{c}$-\frac{6}{xM}[\kt (M(3x-2)+M_X(2-5x))$\\$+x(M+M_X)(M(x-1)+M_X)^2]$ \end{tabular}&\begin{tabular}{c}$\frac{3}{xM^2}(M+M_X)[\kt (M(-4x^2+3x+2)$\\$-M_X(5x+2))+x(M+M_X)(M(x-1)+M_X)^2]$ \end{tabular}\\
        \hline
        $l = 3$ & 0 & $-\frac{3\kt (2\kt +3x(M+M_X)(M(x-1)+M_X))}{xM^2}$\\
        \hline
        $l = 4$ &$\frac{6\kt (M_X-M)}{M}$ &$\frac{3\kt (2\kt +Mx^2(M+M_X))}{xM^2}$\\
        \hline
        $l = 5$ & 0 & 0 \\
        \hline
        $l = 6$ & 0 &0\\
        \hline
        $l = 7$ & 0 & 0 \\
        \hline
        $l = 8$ & 0 & 0 \\
        \hline
    \end{tabular}
\end{table}

\begin{table}[H]
    \centering
    \caption{Coefficients functions of $\Delta H_{3L}^\perp$ for $ i j k = \{221,222\} $.}
    \renewcommand{\arraystretch}{2}
    \begin{tabular}{|c|c|c|}
        \hline
        $C_{ijk}^{[\Delta H_{3L}^\perp], l}$ & $i j k =221$ & $i j k =222$ \\
        \hline
        $l = 1$ & 0& 0 \\
        \hline
        $l = 2$ & $\frac{12\kt }{xM^2}[\kt +M^2(x-1)^2+2MM_X(x-1)x+M_X^2(2x-1)]$ &$-\frac{12\kt (M+M_X)(\kt +xM_X(M(x-1)+M_X))}{xM^3}$\\
        \hline
        $l = 3$ & $\frac{6\kt (2\kt +x(M+M_X)(M(x-1)+M_X))}{xM^2}$ & \begin{tabular}{c} $\frac{3\kt }{2xM^3}[\kt (M(x-6)-3M_X(x+2))+x(M+M_X)$\\$\times(M^2(3x^2-8x+5)-2MM_X x-5M_X^2)]$   \end{tabular}\\
        \hline
        $l = 4$ &$\frac{6\kt (\kt -xM(M(x-1)+M_X))}{xM^2}$ &\begin{tabular}{c} $-\frac{3\kt }{2xM^3}[\kt (M(3x+4)-M_X(x-4))+x(M+M_X)$\\$\times(M^2(x^2-4x+3)-2MM_X-M_X^2)]$   \end{tabular}\\
        \hline
        $l = 5$ & $\frac{3\kt (2\kt +x(M+M_X)(M(x-1)+M_X))}{xM^2}$ & $-\frac{3\kt (M+M_X)(2\kt (x+1)+x(M+M_X)(M(x-1)+M_X))}{2xM^3}$ \\
        \hline
        $l = 6$ & 0 &0\\
        \hline
        $l = 7$ & 0 & 0 \\
        \hline
        $l = 8$ & 0 & 0 \\
        \hline
    \end{tabular}
\end{table}

\subsection{\texorpdfstring{$\Delta H_{3T}$}{DeltaH3T}}

\begin{table}[H]
    \centering
    \caption{Coefficients functions of $\Delta H_{3T}$ for $ i j k = \{111,112\} $.}
    \renewcommand{\arraystretch}{2}
    \begin{tabular}{|c|c|c|}
        \hline
        $C_{ijk}^{[\Delta H_{3T}], l}$ & $i j k = 111$ & $i j k = 112$ \\
        \hline
        $l = 1$ & 0& $\frac{12\kt (\kt -M^2(x-1)^2+M_X^2)}{xM^2}$ \\
        \hline
        $l = 2$ & $\frac{12\kt (5x-1)(M(x-1)+M_X)}{xM}$ &\begin{tabular}{c} $\frac{6\kt }{xM^2}[\kt (x+2)-M^2(x^3+3x^2-5x+1)$\\$+MM_X(1-5x)x+M_X^2(1-4x)]$   \end{tabular}\\
        \hline
        $l = 3$ & $\frac{12\kt (M(x-1)+M_X)}{M}$ & $\frac{3\ktquad (3x-4)}{2xM^2}$\\
        \hline
        $l = 4$ &0 &$\frac{3\kt (\kt (x+4)-4x(M+M_X)(M(x-1)+M_X))}{2xM^2}$\\
        \hline
        $l = 5$ & 0 & 0 \\
        \hline
        $l = 6$ & 0 &0\\
        \hline
        $l = 7$ & 0 & 0 \\
        \hline
        $l = 8$ & 0 & 0 \\
        \hline
    \end{tabular}
\end{table}

\begin{table}[H]
    \centering
    \caption{Coefficients functions of $\Delta H_{3T}$ for $ i j k = \{121,122\} $.}
    \renewcommand{\arraystretch}{2}
    \begin{tabular}{|c|c|c|}
        \hline
        $C_{ijk}^{[\Delta H_{3T}], l}$ & $i j k = 121$ & $i j k = 122$ \\
        \hline
        $l = 1$ & 0& 0 \\
        \hline
        $l = 2$ & $\frac{24\kt (\kt +x(M(x-1)+M_X)^2)}{xM^2}$ &$-\frac{12\kt (\kt (Mx+M_X)+x(M+M_X)(M(x-1)+M_X)^2)}{xM^3}$\\
        \hline
        $l = 3$ & $\frac{12\ktquad}{xM^2}$ & $-\frac{3\ktquad (M(6x^2-8x+6)+3M_X(x+2))}{2xM^3}$\\
        \hline
        $l = 4$ &$\frac{6\kt (\kt (2x+1)+x(M(x-1)+M_X)^2)}{xM^2}$ &\begin{tabular}{c}$\frac{3\kt}{2xM^3}[\kt (2M(x^2-6x+2)-M_X(x+4))$\\$-2x(M+M_X)(M(x-1)+M_X)^2]$ \end{tabular}\\
        \hline
        $l = 5$ & $\frac{6\ktquad }{xM^2}$ & $-\frac{3\ktquad (M(x^2-x+1)+M_X(x+1))}{xM^3}$ \\
        \hline
        $l = 6$ & 0 &0\\
        \hline
        $l = 7$ & 0 & 0 \\
        \hline
        $l = 8$ & 0 & 0 \\
        \hline
    \end{tabular}
\end{table}

\begin{table}[H]
    \centering
    \caption{Coefficients functions of $\Delta H_{3T}$ for $ i j k = \{211,212\} $.}
    \renewcommand{\arraystretch}{2}
    \begin{tabular}{|c|c|c|}
        \hline
        $C_{ijk}^{[\Delta H_{3T}], l}$ & $i j k = 211$ & $i j k = 212$ \\
        \hline
        $l = 1$ & $-\frac{12\kt (\kt -M^2(x-1)^2+M_X^2)}{xM^2}$& 0 \\
        \hline
        $l = 2$ & $-\frac{6\kt (3\kt+x(4M+5M_X)(M(x-1)+M_X))}{xM^2}$ &$\frac{3\kt(M+M_X)[x(M^2(x^2+2x-3)+MM_X(5x-1)+4M_X^2)-\kt (x-1)]}{xM^3}$\\
        \hline
        $l = 3$ & $\frac{3\ktquad}{2M^2}$ & $-\frac{27\ktquad M_X}{8M^3}$\\
        \hline
        $l = 4$ &$-\frac{3\kt(\kt(x+4)+4x(M+M_X)(M(x-1)+M_X))}{2xM^2}$ &$\frac{3\kt}{8xM^3}[\kt(M_X(x+8)-8M(x-1))+8x(M+M_X)^2(M(x-1)+M_X)]$\\
        \hline
        $l = 5$ & 0 & 0 \\
        \hline
        $l = 6$ & 0 &0\\
        \hline
        $l = 7$ & 0 & 0 \\
        \hline
        $l = 8$ & 0 & 0 \\
        \hline
    \end{tabular}
\end{table}

\begin{table}[H]
    \centering
    \caption{Coefficients functions of $\Delta H_{3T}$ for $ i j k = \{221,222\} $.}
    \renewcommand{\arraystretch}{2}
    \begin{tabular}{|c|c|c|}
        \hline
        $C_{ijk}^{[\Delta H_{3T}], l}$ & $i j k =221$ & $i j k =222$ \\
        \hline
        $l = 1$ & 0& 0 \\
        \hline
        $l = 2$ & $-\frac{12\kt [\kt(Mx+M_X)+x(M+M_X)(M(x-1)+M_X)^2]}{xM^3}$ &$\frac{6\kt(M+M_X)}{xM^4}[\kt(M(2x-1)+M_X)+x(M+M_X)(M(x-1)+M_X)^2]$\\
        \hline
        $l = 3$ & $\frac{3\ktquad (M_X-4M)}{2M^3}$ & $\frac{3\ktquad (4\kt+x(8xM^2+4MM_X(3x+1)+3M_X^2))}{8xM^4}$\\
        \hline
        $l = 4$ &\begin{tabular}{c} $-\frac{3\kt}{2xM^3}[\kt(M(8x-2)+M_X(5x+2))$\\$+2x(M+M_X)(M(x-1)+M_X)^2]$ \end{tabular}&\begin{tabular}{c} $-\frac{3\kt}{8xM^4}[4\ktquad-\kt(12M^2(2x-1)-4MM_X(x-7)x$\\$+M_X^2(5x+12))-4x(M+M_X)^2(M(x-1)+M_X)^2]$    \end{tabular}\\
        \hline
        $l = 5$ & $-\frac{3\ktquad}{M^2}$ & $\frac{3\ktquad(M+M_X)(xM+M_X)}{2M^4}$ \\
        \hline
        $l = 6$ & 0 &0\\
        \hline
        $l = 7$ & 0 & 0 \\
        \hline
        $l = 8$ & 0 & 0 \\
        \hline
    \end{tabular}
\end{table}

\subsection{\texorpdfstring{$\Delta H_{3T}^\perp$}{DeltaH3Tperp}}

\begin{table}[H]
    \centering
    \caption{Coefficients functions of $\Delta H_{3T}^\perp$ for $ i j k = \{111,112\} $.}
    \renewcommand{\arraystretch}{2}
    \begin{tabular}{|c|c|c|}
        \hline
        $C_{ijk}^{[\Delta H_{3T}^\perp], l}$ & $i j k = 111$ & $i j k = 112$ \\
        \hline
        $l = 1$ & 0& $\frac{24(\kt-M^2(x-1)^2+M_X^2)}{x}$ \\
        \hline
        $l = 2$ & $\frac{24 M(x-1)(M(x-1)+M_X)}{x}$ &$\frac{12}{x}[\kt(x+2)-M^2(x-1)^2(x+1)-MM_X(x-1)x+M_X^2]$\\
        \hline
        $l = 3$ & 0 & $\frac{3\kt(9x-4)}{x}$\\
        \hline
        $l = 4$ &0 &$\frac{3\kt(4-5x)}{x}$\\
        \hline
        $l = 5$ & 0 & 0 \\
        \hline
        $l = 6$ & 0 &0\\
        \hline
        $l = 7$ & 0 & 0 \\
        \hline
        $l = 8$ & 0 & 0 \\
        \hline
    \end{tabular}
\end{table}

\begin{table}[H]
    \centering
    \caption{Coefficients functions of $\Delta H_{3T}^\perp$ for $ i j k = \{121,122\} $.}
    \renewcommand{\arraystretch}{2}
    \begin{tabular}{|c|c|c|}
        \hline
        $C_{ijk}^{[\Delta H_{3T}^\perp], l}$ & $i j k = 121$ & $i j k = 122$ \\
        \hline
        $l = 1$ & 0& 0 \\
        \hline
        $l = 2$ & $\frac{48\kt(1-x)}{x}$ &$\frac{24\kt M_X(x-1)}{xM}$\\
        \hline
        $l = 3$ & $\frac{24\kt(1-x)}{x}$ & $-\frac{3\kt(6M(x-1)^2+M_X(6-7x))}{xM}$\\
        \hline
        $l = 4$ &$\frac{12\kt(1-x)}{x}$ &$\frac{3\kt(2M(x^2-3x+2)-M_X(x+4))}{xM}$\\
        \hline
        $l = 5$ & $\frac{12\kt(1-x)}{x}$ & $-\frac{6\kt(M(x-1)^2+M_X)}{xM}$ \\
        \hline
        $l = 6$ & 0 &0\\
        \hline
        $l = 7$ & 0 & 0 \\
        \hline
        $l = 8$ & 0 & 0 \\
        \hline
    \end{tabular}
\end{table}

\begin{table}[H]
    \centering
    \caption{Coefficients functions of $\Delta H_{3T}^\perp$ for $ i j k = \{211,212\} $.}
    \renewcommand{\arraystretch}{2}
    \begin{tabular}{|c|c|c|}
        \hline
        $C_{ijk}^{[\Delta H_{3T}^\perp], l}$ & $i j k = 211$ & $i j k = 212$ \\
        \hline
        $l = 1$ & $-\frac{24(\kt-M^2(x-1)^2+M_X^2)}{x}$& 0 \\
        \hline
        $l = 2$ & $-\frac{12}{x}[3\kt+M_X x(M(x-1)+M_X)]$ &$\frac{6(1-x)(M+M_X)(\kt-Mx(M(x-1)+M_X))}{xM}$\\
        \hline
        $l = 3$ & $-3\kt$ & $-\frac{3\kt(8M+15M_X)}{4M}$\\
        \hline
        $l = 4$ &$\frac{3\kt(x-4)}{x}$ &$\frac{3\kt(8M+M_X(7x+8))}{4xM}$\\
        \hline
        $l = 5$ & 0 & 0 \\
        \hline
        $l = 6$ & 0 &0\\
        \hline
        $l = 7$ & 0 & 0 \\
        \hline
        $l = 8$ & 0 & 0 \\
        \hline
    \end{tabular}
\end{table}

\begin{table}[H]
    \centering
    \caption{Coefficients functions of $\Delta H_{3T}^\perp$ for $ i j k = \{221,222\} $.}
    \renewcommand{\arraystretch}{2}
    \begin{tabular}{|c|c|c|}
        \hline
        $C_{ijk}^{[\Delta H_{3T}^\perp], l}$ & $i j k =221$ & $i j k =222$ \\
        \hline
        $l = 1$ & 0& 0 \\
        \hline
        $l = 2$ & $\frac{24\kt M_X(x-1)}{xM}$ &$\frac{12\kt(x-1)(M^2-M_X^2)}{xM^2}$\\
        \hline
        $l = 3$ & $\frac{9\kt M_X}{M}$ & $\frac{3\kt(4\kt+x(8M^2(x-1)+4MM_X(3x-5)-11M_X^2))}{4xM^2}$\\
        \hline
        $l = 4$ &$\frac{3\kt(M_X(3x-2)-2M(x-1))}{xM}$ &$-\frac{3\kt}{4xM^2}[4\kt-12M^2(x-1)+4MM_X(x-3)x+M_X^2(x-12)]$\\
        \hline
        $l = 5$ & $\frac{6\kt M_X}{M}$ & $\frac{3\kt(x-1)(M+M_X)}{M}$ \\
        \hline
        $l = 6$ & 0 &0\\
        \hline
        $l = 7$ & 0 & 0 \\
        \hline
        $l = 8$ & 0 & 0 \\
        \hline
    \end{tabular}
\end{table}

\subsection{\texorpdfstring{Dynamical TMD $\tilde{N}$}{Dynamical TMD N}}

\begin{table}[H]
    \centering
    \caption{Coefficients functions of $\tilde{N}$ for $ i j k = \{111,112\} $.}
    \renewcommand{\arraystretch}{1.5}
    \begin{tabular}{|c|c|c|}
        \hline
        $C_{ijk}^{[\tilde{N}], l}$ & $i j k = 111$ & $i j k = 112$ \\
        \hline
        $l = 1$ & 0& 0 \\
        \hline
        $l = 2$ & 0 &0\\
        \hline
        $l = 3$ & $\frac{12\kt (1-x)(M(x-1)+M_X)}{M}$ & \begin{tabular}{c} $\frac{3\kt}{4M^2(x-1)x} [\kt(x^3-10x^2+24x-16)$\\$+x^2(M^2(x-1)^2(3x-2)+8MM_X(x-1)^2+M_X^2(7x-8))]$    \end{tabular}\\
        \hline
        $l = 4$ &$\frac{12\kt(x-1)(M(x-1)+M_X)}{M}$ &\begin{tabular}{c} $-\frac{3\kt}{4M^2(x-1)x} [\kt(x^3-10x^2+24x-16)$\\$+x^2(M^2(x-1)^2(3x-2)+8MM_X(x-1)^2+M_X^2(7x-8))]$    \end{tabular}\\
        \hline
        $l = 5$ & 0 & $\frac{3\ktquad (x^2-3x+4)}{xM^2}$ \\
        \hline
        $l = 6$ & 0 &$-\frac{3\ktquad x}{4M^2}$\\
        \hline
        $l = 7$ & 0 & $\frac{3\ktquad x}{4M^2}$ \\
        \hline
        $l = 8$ & 0 & $-\frac{3\ktquad(x^2-3x+4)}{xM^2}$ \\
        \hline
        $l = 9$ & 0 &0\\
        \hline
        $l = 10$ & 0 & 0 \\
        \hline
        $l = 11$ & 0 & 0 \\
        \hline
    \end{tabular}
\end{table}

\begin{table}[H]
    \centering
    \caption{Coefficients functions of $\tilde{N}$ for $ i j k = \{121,122\} $.}
    \renewcommand{\arraystretch}{1.5}
    \begin{tabular}{|c|c|c|}
        \hline
        $C_{ijk}^{[\tilde{N}], l}$ & $i j k = 121$ & $i j k = 122$ \\
        \hline
        $l = 1$ & 0& 0 \\
        \hline
        $l = 2$ & 0 &0\\
        \hline
        $l = 3$ & 0 & \begin{tabular}{c}$\frac{3\kt}{16M^3(1-x)}[\kt(2M(3x^2-7x+4)+M_X(-18x^2+49x-32))$\\$+M_X x(M^2(x-1)^2(2x+3)+M_X^2(2x-3))]$   \end{tabular}\\
        \hline
        $l = 4$ &0 &\begin{tabular}{c}$\frac{3\kt}{16M^3(x-1)}[\kt(2M(3x^2-7x+4)+M_X(-18x^2+49x-32))$\\$+M_X x(M^2(x-1)^2(2x+3)+M_X^2(2x-3))]$   \end{tabular}\\
        \hline
        $l = 5$ & $-\frac{3\ktquad(x^2-6x+4)}{xM^2}$ & $-\frac{3\ktquad (2M(x^3-7x^2+10x-4)+M_X(4x^2-13x+8))}{8M^3(1-x)}$ \\
        \hline
        $l = 6$ &$-\frac{3\ktquad x}{2M^2}$ & $-\frac{3\ktquad (2M(x-1)x^2+M_X(8x^2-23x+16))}{16M^3(1-x)}$\\
        \hline
        $l = 7$ & $\frac{3\ktquad x}{2M^2}$ & $\frac{3\ktquad (2M(x-1)x^2+M_X(8x^2-23x+16))}{16M^3(1-x)}$ \\
        \hline
        $l = 8$ & $\frac{3\ktquad(x^2-6x+4)}{xM^2}$ & $\frac{3\ktquad (2M(x^3-7x^2+10x-4)+M_X(4x^2-13x+8))}{8M^3(1-x)}$ \\
        \hline
        $l = 9$ & 0 &0\\
        \hline
        $l = 10$ & 0 & 0 \\
        \hline
        $l = 11$ & 0 & 0 \\
        \hline
    \end{tabular}
\end{table}

\begin{table}[H]
    \centering
    \caption{Coefficients functions of $\tilde{N}$ for $ i j k = \{211,212\} $.}
    \renewcommand{\arraystretch}{1.5}
    \begin{tabular}{|c|c|c|}
        \hline
        $C_{ijk}^{[\tilde{N}], l}$ & $i j k = 211$ & $i j k = 212$ \\
        \hline
        $l = 1$ & 0& 0 \\
        \hline
        $l = 2$ & 0 &0\\
        \hline
        $l = 3$ & \begin{tabular}{c}$\frac{3\kt}{4xM^2} [\kt(5x^2+8x-16)+x(M^2(-x^3+8x^2-15x+8)$\\$+8MM_X(x-1)x+M_X^2(7x-8))]$   \end{tabular} & \begin{tabular}{c} $-\frac{3\kt}{16M^3(x-1)}[\kt(M(12x^2-17x+4)+4M_X(2x^2-x-1))$\\$+x(M^3(x-1)^2(6x-7)+4M^2 M_X(x-1)^2(x+3)$\\$+MM_X^2(16x^2-24x+7)+12M_X^3(x-1))]$     \end{tabular}\\
        \hline
        $l = 4$ &\begin{tabular}{c}$-\frac{3\kt}{4xM^2} [\kt(5x^2+8x-16)+x(M^2(-x^3+8x^2-15x+8)$\\$+8MM_X(x-1)x+M_X^2(7x-8))]$   \end{tabular} &\begin{tabular}{c} $\frac{3\kt}{16M^3(x-1)}[\kt(M(12x^2-17x+4)+4M_X(2x^2-x-1))$\\$+x(M^3(x-1)^2(6x-7)+4M^2 M_X(x-1)^2(x+3)$\\$+MM_X^2(16x^2-24x+7)+12M_X^3(x-1))]$     \end{tabular}\\
        \hline
        $l = 5$ & 0 & $-\frac{3\ktquad (x-2)(M_X(3x-4)-2M(x-1))}{8M^3(x-1)}$ \\
        \hline
        $l = 6$ & 0 &$-\frac{3\ktquad M_X x^2}{16 M^3 (1-x)}$\\
        \hline
        $l = 7$ & 0 & $\frac{3\ktquad M_X x^2}{16 M^3 (1-x)}$ \\
        \hline
        $l = 8$ & 0 & $\frac{3\ktquad (x-2)(M_X(3x-4)-2M(x-1))}{8M^3(x-1)}$ \\
        \hline
        $l = 9$ & 0 &0\\
        \hline
        $l = 10$ & 0 & 0 \\
        \hline
        $l = 11$ & 0 & 0 \\
        \hline
    \end{tabular}
\end{table}

\begin{table}[H]
    \centering
    \caption{Coefficients functions of $\tilde{N}$ for $ i j k = \{221,222\} $.}
    \renewcommand{\arraystretch}{1.5}
    \begin{tabular}{|c|c|c|}
        \hline
        $C_{ijk}^{[\tilde{N}], l}$ & $i j k =221$ & $i j k =222$ \\
        \hline
        $l = 1$ & 0& 0 \\
        \hline
        $l = 2$ & 0 &0\\
        \hline
        $l = 3$ & \begin{tabular}{c} $\frac{3\kt}{16M^3(x-1)}[\kt(2M(3x^2-7x+4)+M_X(-18x^2+49x-32))$\\$+M_X x(M^2(x-1)^2(2x+3)+M_X^2(2x-3))]$     \end{tabular} & 0\\
        \hline
        $l = 4$ &\begin{tabular}{c} $-\frac{3\kt}{16M^3(x-1)}[\kt(2M(3x^2-7x+4)+M_X(-18x^2+49x-32))$\\$+M_X x(M^2(x-1)^2(2x+3)+M_X^2(2x-3))]$     \end{tabular} &\begin{tabular}{c}$\frac{3\ktquad}{32M^4(x-1)x}[2\kt(2x^3-8x^2+21x-16)$\\$+x(2M^2(2x^4-9x^3+16x^2-14x+5)$\\$+MM_X x(6x^2-17x+8)+2M_X^2(6x^2-6x-5))]$ \end{tabular}\\
        \hline
        $l = 5$ & $-\frac{3\ktquad (M(2x^3-13x^2+17x-6)+2M_X(3x^2-8x+4))}{8M^3(1-x)}$ & \begin{tabular}{c}$\frac{3\ktquad}{32M^4(x-1)x}[2\kt(2x^3-8x^2+21x-16)$\\$+x(2M^2(2x^4-9x^3+16x^2-14x+5)$\\$+MM_X x(6x^2-17x+8)+2M_X^2(6x^2-6x-5))]$ \end{tabular} \\
        \hline
        $l = 6$ &$-\frac{3\ktquad x^2(M(x-1)+M_X)}{8M^3(1-x)}$ & \begin{tabular}{c}$-\frac{3\ktquad}{64M^4(x-1)x}[4\kt(6x^3+6x^2-33x+16)$\\$+x(2M^2(8x^3-25x^2+27x-10)$\\$+MM_X x(8x^2-21x+16)+10M_X^2(x^2-3x+2))]$ \end{tabular}\\
        \hline
        $l = 7$ & $\frac{3\ktquad x^2(M(x-1)+M_X)}{8M^3(1-x)}$ & \begin{tabular}{c}$\frac{3\ktquad}{64M^4(x-1)x}[4\kt(6x^3+6x^2-33x+16)$\\$+x(2M^2(8x^3-25x^2+27x-10)$\\$+MM_X x(8x^2-21x+16)+10M_X^2(x^2-3x+2))]$ \end{tabular}\\
        \hline
        $l = 8$ & $\frac{3\ktquad (M(2x^3-13x^2+17x-6)+2M_X(3x^2-8x+4))}{8M^3(1-x)}$ & \begin{tabular}{c}$\frac{3\ktquad}{32M^4(x-1)x}[2\kt(2x^3-8x^2+21x-16)$\\$+x(2M^2(2x^4-9x^3+16x^2-14x+5)$\\$+MM_X x(6x^2-17x+8)+2M_X^2(6x^2-6x-5))]$ \end{tabular}\\
        \hline
        $l = 9$ & 0 &$\frac{3\bm{k}_T^6 (4x^2-x-6)}{16M^4(1-x)}$\\
        \hline
        $l = 10$ & 0 & $\frac{3\bm{k}_T^6(2x^2-3x-2)}{32M^4(x-1)}$ \\
        \hline
        $l = 11$ & 0 & $-\frac{3\bm{k}_T^6(2x^2-3x-2)}{32M^4(x-1)}$ \\
        \hline
    \end{tabular}
\end{table}

\end{widetext}


\begin{thebibliography}{99}

\bibitem{Efremov:1981sh}
A.~V.~Efremov and O.~V.~Teryaev,
Sov. J. Nucl. Phys. \textbf{36}, 140 (1982)
JINR-P2-81-485.

\bibitem{Qiu:1991wg}

J.~w.~Qiu and G.~F.~Sterman,
Nucl. Phys. B \textbf{378}, 52-78 (1992)
doi:10.1016/0550-3213(92)90003-T

\bibitem{Qiu:1998ia}
J.~w.~Qiu and G.~F.~Sterman,
Phys. Rev. D \textbf{59}, 014004 (1999)
doi:10.1103/PhysRevD.59.014004
[arXiv:hep-ph/9806356 [hep-ph]].

\bibitem{Jaffe:1989xx}
R.~L.~Jaffe,
Comments Nucl. Part. Phys. \textbf{19} (1990) no.5, 239-257
MIT-CTP-1798.

\bibitem{Belitsky:2000pb}
A.~V.~Belitsky, X.~D.~Ji, W.~Lu and J.~Osborne,
Phys. Rev. D \textbf{63} (2001), 094012
doi:10.1103/PhysRevD.63.094012
[arXiv:hep-ph/0007305 [hep-ph]].

\bibitem{Kanazawa:2000hz}
Y.~Kanazawa and Y.~Koike,
Phys. Lett. B \textbf{478} (2000), 121-126
doi:10.1016/S0370-2693(00)00261-6
[arXiv:hep-ph/0001021 [hep-ph]].

\bibitem{Kanazawa:2000kp}
Y.~Kanazawa and Y.~Koike,
Phys. Lett. B \textbf{490} (2000), 99-105
doi:10.1016/S0370-2693(00)00969-2
[arXiv:hep-ph/0007272 [hep-ph]].

\bibitem{Ji:2006vf}
X.~Ji, J.~w.~Qiu, W.~Vogelsang and F.~Yuan,
Phys. Rev. D \textbf{73} (2006), 094017
doi:10.1103/PhysRevD.73.094017
[arXiv:hep-ph/0604023 [hep-ph]].

\bibitem{Kouvaris:2006zy}
C.~Kouvaris, J.~W.~Qiu, W.~Vogelsang and F.~Yuan,
Phys. Rev. D \textbf{74} (2006), 114013
doi:10.1103/PhysRevD.74.114013
[arXiv:hep-ph/0609238 [hep-ph]].

\bibitem{Koike:2007rq}
Y.~Koike and K.~Tanaka,
Phys. Rev. D \textbf{76} (2007), 011502
doi:10.1103/PhysRevD.76.011502
[arXiv:hep-ph/0703169 [hep-ph]].

\bibitem{Koike:2009ge}
Y.~Koike and T.~Tomita,
Phys. Lett. B \textbf{675} (2009), 181-189
doi:10.1016/j.physletb.2009.04.017
[arXiv:0903.1923 [hep-ph]].

\bibitem{Vogelsang:2009pj}
W.~Vogelsang and F.~Yuan,
Phys. Rev. D \textbf{79} (2009), 094010
doi:10.1103/PhysRevD.79.094010
[arXiv:0904.0410 [hep-ph]].

\bibitem{Koike:2011mb}
Y.~Koike and S.~Yoshida,
Phys. Rev. D \textbf{84} (2011), 014026
doi:10.1103/PhysRevD.84.014026
[arXiv:1104.3943 [hep-ph]].

\bibitem{Koike:2011nx}
Y.~Koike and S.~Yoshida,
Phys. Rev. D \textbf{85} (2012), 034030
doi:10.1103/PhysRevD.85.034030
[arXiv:1112.1161 [hep-ph]].

\bibitem{Kang:2010zzb}
Z.~B.~Kang, F.~Yuan and J.~Zhou,
Phys. Lett. B \textbf{691} (2010), 243-248
doi:10.1016/j.physletb.2010.07.003
[arXiv:1002.0399 [hep-ph]].

\bibitem{Kanazawa:2011er}
K.~Kanazawa and Y.~Koike,
Phys. Lett. B \textbf{701} (2011), 576-580
doi:10.1016/j.physletb.2011.06.021
[arXiv:1105.1036 [hep-ph]].

\bibitem{Metz:2012ct}
A.~Metz and D.~Pitonyak,
Phys. Lett. B \textbf{723} (2013), 365-370
[erratum: Phys. Lett. B \textbf{762} (2016), 549-549]
doi:10.1016/j.physletb.2013.05.043
[arXiv:1212.5037 [hep-ph]].

\bibitem{Beppu:2013uda}
H.~Beppu, K.~Kanazawa, Y.~Koike and S.~Yoshida,
Phys. Rev. D \textbf{89} (2014) no.3, 034029
doi:10.1103/PhysRevD.89.034029
[arXiv:1312.6862 [hep-ph]].

\bibitem{Kanazawa:2014nea}
K.~Kanazawa, Y.~Koike, A.~Metz and D.~Pitonyak,
Phys. Rev. D \textbf{91} (2015) no.1, 014013
doi:10.1103/PhysRevD.91.014013
[arXiv:1410.3448 [hep-ph]].

\bibitem{Kang:2011ni}
Z.~B.~Kang and F.~Yuan,
Phys. Rev. D \textbf{84} (2011), 034019
doi:10.1103/PhysRevD.84.034019
[arXiv:1106.1375 [hep-ph]].

\bibitem{Hatta:2016wjz}
Y.~Hatta, B.~W.~Xiao, S.~Yoshida and F.~Yuan,
Phys. Rev. D \textbf{94} (2016) no.5, 054013
doi:10.1103/PhysRevD.94.054013
[arXiv:1606.08640 [hep-ph]].

\bibitem{Hatta:2016khv}
Y.~Hatta, B.~W.~Xiao, S.~Yoshida and F.~Yuan,
Phys. Rev. D \textbf{95} (2017) no.1, 014008
doi:10.1103/PhysRevD.95.014008
[arXiv:1611.04746 [hep-ph]].

\bibitem{Benic:2018moa}
S.~Beni{\'c} and Y.~Hatta,
Phys. Rev. D \textbf{98} (2018) no.9, 094025
doi:10.1103/PhysRevD.98.094025
[arXiv:1806.10901 [hep-ph]].

\bibitem{Benic:2018amn}
S.~Beni{\'c} and Y.~Hatta,
Phys. Rev. D \textbf{99} (2019) no.9, 094012
doi:10.1103/PhysRevD.99.094012
[arXiv:1811.10589 [hep-ph]].

\bibitem{Ji:2006br}
X.~Ji, J.~W.~Qiu, W.~Vogelsang and F.~Yuan,
Phys. Lett. B \textbf{638} (2006), 178-186
doi:10.1016/j.physletb.2006.05.044
[arXiv:hep-ph/0604128 [hep-ph]].

\bibitem{Eguchi:2006qz}
H.~Eguchi, Y.~Koike and K.~Tanaka,
Nucl. Phys. B \textbf{752} (2006), 1-17
doi:10.1016/j.nuclphysb.2006.05.036
[arXiv:hep-ph/0604003 [hep-ph]].

\bibitem{Eguchi:2006mc}
H.~Eguchi, Y.~Koike and K.~Tanaka,
Nucl. Phys. B \textbf{763} (2007), 198-227
doi:10.1016/j.nuclphysb.2006.11.016
[arXiv:hep-ph/0610314 [hep-ph]].

\bibitem{Yuan:2009dw}
F.~Yuan and J.~Zhou,
Phys. Rev. Lett. \textbf{103} (2009), 052001
doi:10.1103/PhysRevLett.103.052001
[arXiv:0903.4680 [hep-ph]].

\bibitem{Beppu:2010qn}
H.~Beppu, Y.~Koike, K.~Tanaka and S.~Yoshida,
Phys. Rev. D \textbf{82} (2010), 054005
doi:10.1103/PhysRevD.82.054005
[arXiv:1007.2034 [hep-ph]].

\bibitem{Koike:2011ns}
Y.~Koike, K.~Tanaka and S.~Yoshida,
Phys. Rev. D \textbf{83} (2011), 114014
doi:10.1103/PhysRevD.83.114014
[arXiv:1104.0798 [hep-ph]].

\bibitem{Kanazawa:2013uia}
K.~Kanazawa and Y.~Koike,
Phys. Rev. D \textbf{88} (2013), 074022
doi:10.1103/PhysRevD.88.074022
[arXiv:1309.1215 [hep-ph]].

\bibitem{Yoshida:2016tfh}
S.~Yoshida,
Phys. Rev. D \textbf{93} (2016) no.5, 054048
doi:10.1103/PhysRevD.93.054048
[arXiv:1601.07737 [hep-ph]].

\bibitem{Xing:2019ovj}
H.~Xing and S.~Yoshida,
Phys. Rev. D \textbf{100} (2019) no.5, 054024
doi:10.1103/PhysRevD.100.054024
[arXiv:1904.02287 [hep-ph]].

\bibitem{Benic:2019zvg}
S.~Benic, Y.~Hatta, H.~n.~Li and D.~J.~Yang,
Phys. Rev. D \textbf{100} (2019) no.9, 094027
doi:10.1103/PhysRevD.100.094027
[arXiv:1909.10684 [hep-ph]].

\bibitem{Kanazawa:2000cx}
Y.~Kanazawa and Y.~Koike,
Phys. Rev. D \textbf{64} (2001), 034019
doi:10.1103/PhysRevD.64.034019
[arXiv:hep-ph/0012225 [hep-ph]].

\bibitem{Zhou:2008fb}
J.~Zhou, F.~Yuan and Z.~T.~Liang,
Phys. Rev. D \textbf{78} (2008), 114008
doi:10.1103/PhysRevD.78.114008
[arXiv:0808.3629 [hep-ph]].

\bibitem{Koike:2015zya}
Y.~Koike, K.~Yabe and S.~Yoshida,
Phys. Rev. D \textbf{92} (2015) no.9, 094011
doi:10.1103/PhysRevD.92.094011
[arXiv:1509.06830 [hep-ph]].

\bibitem{Koike:2017fxr}
Y.~Koike, A.~Metz, D.~Pitonyak, K.~Yabe and S.~Yoshida,
Phys. Rev. D \textbf{95} (2017) no.11, 114013
doi:10.1103/PhysRevD.95.114013
[arXiv:1703.09399 [hep-ph]].

\bibitem{Yabe:2019awq}
K.~Yabe, Y.~Koike, A.~Metz, D.~Pitonyak and S.~Yoshida,
JPS Conf. Proc. \textbf{26} (2019), 021016
doi:10.7566/JPSCP.26.021016

\bibitem{Kenta:2019bxd}
Y.~Kenta, Y.~Koike, A.~Metz, D.~Pitonyak and S.~Yoshida,
PoS \textbf{SPIN2018} (2019), 192
doi:10.22323/1.346.0192

\bibitem{Gamberg:2018fwy}
L.~Gamberg, Z.~B.~Kang, D.~Pitonyak, M.~Schlegel and S.~Yoshida,
JHEP \textbf{01} (2019), 111
doi:10.1007/JHEP01(2019)111
[arXiv:1810.08645 [hep-ph]].

\bibitem{Jaffe:1991ra}
R.~L.~Jaffe and X.~D.~Ji,
Nucl. Phys. B \textbf{375} (1992), 527-560
doi:10.1016/0550-3213(92)90110-W

\bibitem{Liang:2012rb}
Z.~T.~Liang, A.~Metz, D.~Pitonyak, A.~Sch{\"a}fer, Y.~K.~Song and J.~Zhou,
Phys. Lett. B \textbf{712} (2012), 235-239
doi:10.1016/j.physletb.2012.04.072
[arXiv:1203.3956 [hep-ph]].

\bibitem{Hatta:2013wsa}
Y.~Hatta, K.~Kanazawa and S.~Yoshida,
Phys. Rev. D \textbf{88} (2013) no.1, 014037
doi:10.1103/PhysRevD.88.014037
[arXiv:1305.7001 [hep-ph]].

\bibitem{Koike:2015yza}
Y.~Koike, D.~Pitonyak, Y.~Takagi and S.~Yoshida,
Phys. Lett. B \textbf{752} (2016), 95-101
doi:10.1016/j.physletb.2015.11.014
[arXiv:1508.06499 [hep-ph]].

\bibitem{Koike:2016ura}
Y.~Koike, D.~Pitonyak and S.~Yoshida,
Phys. Lett. B \textbf{759} (2016), 75-81
doi:10.1016/j.physletb.2016.05.043
[arXiv:1603.07908 [hep-ph]].

\bibitem{Kanazawa:2015ajw}
K.~Kanazawa, Y.~Koike, A.~Metz, D.~Pitonyak and M.~Schlegel,
Phys. Rev. D \textbf{93} (2016) no.5, 054024
doi:10.1103/PhysRevD.93.054024
[arXiv:1512.07233 [hep-ph]].

\bibitem{Bacchetta:2004zf}
A.~Bacchetta, P.~J.~Mulders and F.~Pijlman,
Phys. Lett. B \textbf{595} (2004), 309-317
doi:10.1016/j.physletb.2004.06.052
[arXiv:hep-ph/0405154 [hep-ph]].

\bibitem{Bacchetta:2006tn}
A.~Bacchetta, M.~Diehl, K.~Goeke, A.~Metz, P.~J.~Mulders and M.~Schlegel,
JHEP \textbf{02} (2007), 093
doi:10.1088/1126-6708/2007/02/093
[arXiv:hep-ph/0611265 [hep-ph]].

\bibitem{Gamberg:2022lju}
L.~Gamberg, Z.~B.~Kang, D.~Y.~Shao, J.~Terry and F.~Zhao,
[arXiv:2211.13209 [hep-ph]].

\bibitem{Ebert:2021jhy}
M.~A.~Ebert, A.~Gao and I.~W.~Stewart,
JHEP \textbf{06} (2022), 007
[erratum: JHEP \textbf{07} (2023), 096]
doi:10.1007/JHEP06(2022)007
[arXiv:2112.07680 [hep-ph]].

\bibitem{Koike:2019zxc}
Y.~Koike, K.~Yabe and S.~Yoshida,
Phys. Rev. D \textbf{101} (2020) no.5, 054017
doi:10.1103/PhysRevD.101.054017
[arXiv:1912.11199 [hep-ph]].

\bibitem{Mulders:2000sh}
P.~J.~Mulders and J.~Rodrigues,
Phys. Rev. D \textbf{63} (2001), 094021
doi:10.1103/PhysRevD.63.094021
[arXiv:hep-ph/0009343 [hep-ph]].

\bibitem{Boer:2016fqd}
D.~Boer, P.~J.~Mulders, C.~Pisano and J.~Zhou,
JHEP \textbf{08} (2016), 001
doi:10.1007/JHEP08(2016)001
[arXiv:1605.07934 [hep-ph]].

\bibitem{Bomhof:2006dp}
C.~J.~Bomhof, P.~J.~Mulders and F.~Pijlman,
Eur. Phys. J. C \textbf{47} (2006), 147-162
doi:10.1140/epjc/s2006-02554-2
[arXiv:hep-ph/0601171 [hep-ph]].

\bibitem{Buffing:2013kca}
M.~G.~A.~Buffing, A.~Mukherjee and P.~J.~Mulders,
Phys. Rev. D \textbf{88} (2013), 054027
doi:10.1103/PhysRevD.88.054027
[arXiv:1306.5897 [hep-ph]].

\bibitem{Goeke:2006ef}
K.~Goeke, S.~Meissner, A.~Metz and M.~Schlegel,
Phys. Lett. B \textbf{637} (2006), 241-244
doi:10.1016/j.physletb.2006.05.004
[arXiv:hep-ph/0601133 [hep-ph]].

\bibitem{Collins:2011zzd}
J.~Collins,
Camb. Monogr. Part. Phys. Nucl. Phys. Cosmol. \textbf{32} (2011), 1-624
Cambridge University Press, 2011,
ISBN 978-1-009-40184-5, 978-1-009-40183-8, 978-1-009-40182-1
doi:10.1017/9781009401845

\bibitem{Bacchetta:2020vty}
A.~Bacchetta, F.~G.~Celiberto, M.~Radici and P.~Taels,
Eur. Phys. J. C \textbf{80} (2020) no.8, 733
doi:10.1140/epjc/s10052-020-8327-6
[arXiv:2005.02288 [hep-ph]].

\bibitem{Bacchetta:2024fci}
A.~Bacchetta, F.~G.~Celiberto and M.~Radici,
Eur. Phys. J. C \textbf{84} (2024) no.6, 576
doi:10.1140/epjc/s10052-024-12927-y
[arXiv:2402.17556 [hep-ph]].

\bibitem{Ji:1992eu}
X.~D.~Ji,
Phys. Lett. B \textbf{289} (1992), 137-142
doi:10.1016/0370-2693(92)91375-J

\bibitem{Lu:2015wja}
Z.~Lu and I.~Schmidt,
Phys. Lett. B \textbf{747} (2015), 357-364
doi:10.1016/j.physletb.2015.06.011
[arXiv:1501.04379 [hep-ph]].

\bibitem{Yang:2016mxl}
Y.~Yang, Z.~Lu and I.~Schmidt,
Phys. Lett. B \textbf{761} (2016), 333-339
doi:10.1016/j.physletb.2016.08.053
[arXiv:1607.01638 [hep-ph]].



\end{thebibliography}
\end{document}